\documentclass{article}
\usepackage{savesym}
\usepackage{amsmath}
\usepackage{cases}
\usepackage{amssymb}
\usepackage[T1]{fontenc}
\usepackage[ansinew]{inputenc}
\usepackage{enumitem}
\usepackage{graphicx}
\usepackage{xspace}
\usepackage{subfigure}
\usepackage{color}
\usepackage[cc]{titlepic}
\usepackage{tabularx}
\usepackage{lipsum,appendix}
\usepackage{caption}
\usepackage{titlesec}
\usepackage{url}
\usepackage{mwe}
\usepackage[textfont=it]{caption}
\usepackage{hyperref}
\hypersetup{colorlinks   = true,citecolor    = blue}
\usepackage{natbib}

\titleformat{\section}[runin]
{\normalfont\bfseries}{\thesection.}{0.5em}{}
\titleformat{\subsection}[runin]
{\normalfont\itshape}{\thesubsection.}{0.5em}{}

\newcommand{\diag}{\mathop{\rm diag}\nolimits}
\newcommand{\argmin}{\mathop{\rm argmin}\limits}
\mathchardef\given="626A
\newcommand{\thalf}{{\textstyle\frac12}}

\title{{\Large\bf An empirical Bayes approach to network recovery using external knowledge}
	\footnote{The research leading to these results has received funding from the European Research Council under ERC Grant Agreement 320637.}
}
\author{ %
	Gino B. Kpogbezan$^1$, Aad W. van der Vaart$^1$, 
	Wessel N. van Wieringen$^{2,3}$,\\
	Gwena\"el G.R. Leday$^4$, Mark A. van de Wiel$^{2,3}$ \\
	\\
	\small	$^{1}$Mathematical Institute, Leiden University\\
	\small	$^{2}$Department of Mathematics, VU University\\
	\small	$^{3}$Department of Epidemiology and Biostatistics,\\
	\small VU University Medical Center\\
	\small	$^{4}$MRC Biostatistics Unit, Cambridge Institute of Public Health
}
\date{}

\begin{document}
	
	\maketitle
	
	\begin{abstract}
		Reconstruction of a high-dimensional network may benefit substantially from
		the inclusion of prior knowledge on the network topology. In the case of gene interaction networks such knowledge may come
		for instance from pathway repositories like KEGG, or be inferred from data of a pilot study. The Bayesian framework
		provides a natural means of including such prior knowledge.  Based on a Bayesian Simultaneous Equation
		Model, we develop an appealing empircal Bayes procedure which automatically assesses the relevance of the used prior knowledge. We use variational Bayes method for posterior densities approximation and compare its accuracy with that of Gibbs sampling strategy.
		Our method is computationally fast, and can outperform known competitors. 
		In a simulation study we show that accurate prior data can greatly improve the
		reconstruction of the network, but need not harm the reconstruction if wrong. We demonstrate the benefits of the method
		in an analysis of gene expression data from GEO. In particular, the edges of the recovered network have superior reproducibility (compared to that of competitors) over resampled versions of the data.
	\end{abstract}

\section{Introduction}
Many areas of the quantitative sciences have witnessed a data deluge in recent years. This is due to
an increased capacity of measuring and storing data in combination with a reduction in costs of
acquiring this data. For instance, in the medical field high-throughput platforms yield measurements
of many molecular aspects (e.g. gene expression) of the cell. As many as $20000$ genes of a single
patient can be characterized simultaneously. However, although the costs of such techniques have
gone down over the years, the number of patients $n$ in a typical clinical study is still small
compared to the number of variables $p$ measured. Reliable analysis of data of such a ``$n\ll p$''
study is difficult.  In this paper we try to solve the problem of few replicate measurements by
incorporating external (or ``prior'') data in the analysis.

High-dimensional modelling based on a small data set is particularly challenging in studies of
relationships between variables. Already the number of binary relationships between even a modest
number of genes is high. However, to some extent these relationships may be known from the vast body
of medical literature available. For instance, the current beliefs on interactions among genes is
condensed in repositories like KEGG and Reactome. Although such information may not be reliable, or
be only partially relevant for the case at hand, its flexible inclusion may help the analysis of
high-dimensional data. Methodology that exploits such prior information may accelerate our
understanding of complex systems like the cell.

The cohesion of variables constituting a complex system is often represented by a network, also
referred to as a \emph{graph}. A graph $\mathcal{G}$ consists of a pair $(\mathcal{I}, \mathcal{E})$ where
$\mathcal{I} = \{1,...,p\}$ is a set of indices representing nodes (the variables of the
system) and $\mathcal{E}$ is the set of edges (relations between the variables) in
$\mathcal{I}\times\mathcal{I}$. An edge can be operationalized in many ways, we concentrate on it
representing conditional independence between the node pair it connects. More formally, a pair
$(i_1, i_2) \in \mathcal{E}$ if and only if random variables represented by nodes $i_1$ and $i_2$
are conditionally \emph{de}pendent, given all remaining nodes in $\mathcal{I}$. All pairs of nodes of
$\mathcal{I}$ not in $\mathcal{E}$ are conditionally independent given the remaining nodes. Graphs
endowed with this operationalization of the edges are referred to as conditional independence graphs (Whittaker, 1990).

Conditional independence graphs are learned from data by graphical models. Graphical models specify
how data are generated obeying the relations among the variables as specified by a conditional
independence graph. A Gaussian Graphical Model (GGM) assumes data are drawn from a multivariate
normal distribution: 
\begin{equation}
\label{EqData}
 Y^j \sim^\text{iid}\ \text{N}( 0 , \Omega^{-1}_p), \qquad j\in \{1,...,n\}.
\end{equation}

Here $Y^j$ is a $p$-dimensional random vector comprising the $p$ random variables
$Y^j_1, \ldots,Y^j_p$ corresponding to the nodes of $\mathcal{I}$ and $\Omega_p^{-1}$ is a
non-singular $(p\times p)$-dimensional covariance matrix. The matrix $\Omega_p$, as opposed to its inverse, is referred
to as the \emph{precision matrix}. For a GGM the edge set $\mathcal{E}$ of
the underlying conditional independence graph corresponds to the nonzero elements of $\Omega_p$ (Lauritzen, 1996).
Hence, to reconstruct the conditional independence graph it suffices to
determine the support of this matrix.

Reconstruction of the conditional independence graph may concentrate on the direct estimation of the
precision matrix. Here we choose a different line of attack. This exploits an equivalence between
Gaussian graphical models and Simultaneous Equations Models (SEMs), which we introduce first before
pointing out the equivalence. We treat SEMs as a system of regression equations, with each equation
modelling the conditional distribution of a node given the other nodes. 
If we collect all observations on node $i\in\mathcal{I}$ in a vector $Y_i:=(Y_i^1,\ldots, Y_i^n)^T$,  then
we can write:
\begin{equation}
\label{EqSEM}
 Y_i=X_i\beta_{i}+\epsilon_i, \quad i\in \mathcal{I}, 
\end{equation}
where $X_i$ is the $n\times (p-1)$-matrix with columns the observations of the $p-1$ nodes different
from $i$, i.e.\ $X_i=[Y_1,Y_2,...,Y_{i-1},Y_{i+1},...,Y_p]$ (where the square brackets mean
``combine the vectors in a matrix''). The error vector $\epsilon_i$ is defined by the equation, and
possesses a multivariate Gaussian distribution $\text{N}(0,\sigma_i^2\bf{I_n})$ under the GGM.  (The
covariances between the errors of different equations are in general non-zero, but are left
unspecified.) The equivalence between the thus formulated SEM and the GGM as specified above stems
from the one-to-one relationship between the regression parameters of the SEM and the elements of the
GGM's precision matrix (confer e.g. Lauritzen (1996)):
$\beta_{i,r}=-\omega_{ii}^{-1}\omega_{ir}$. In particular, (non)zero entries in the $i$-th row
vector of the precision matrix $\Omega_p$ correspond to the (non)zero coefficients of $\beta_i$. The
problem of identifying (non)zero entries in $\Omega_p$ can therefore be cast as a variable selection
problem in the $p$ regression models \eqref{EqSEM}. Lasso regression 
(Tibshirani, 1996) may be
put forward for this purpose (as is done in Meinshausen and B\"uhlmann (2006)), but
other variable selection methods have also been employed.  The slight embarassment that every
partial correlation appears in two regression equations is usually resolved by post-symmetrization
through application of the `AND'-rule: an edge $(i,j) \in \mathcal{E}$ if and only if
$\beta_{i,j} \neq 0$ and $\beta_{j,i} \neq 0$ (Meinshausen and B\"uhlmann, 2006). In combination with this
`AND'-rule, graph structure recovery based on model \eqref{EqSEM} performs well and is widely used in
practice.

Previously, we proposed a Bayesian formulation of the SEM (Leday et al., 2015). In this
Bayesian SEM (henceforth BSEM) the structural model \eqref{EqSEM} is endowed with the following prior:
\begin{equation}
\label{EqBSEM}
\begin{array}{ccc}
 \epsilon_i\given \sigma^2_i ,\tau_i^2\sim \text{N}(0_n, \sigma_i^2{\bf{I}}_n), \\
 \beta_i\given \sigma_i^2,\tau^{2}_i \sim \text{N}(0_s,\sigma_i^2\tau^{-2}_i{\bf{I}}_s),\\
 \tau^{2}_i \sim \Gamma(a_1, b_1),\\
 \sigma^{-2}_i \sim \Gamma(a_2, b_2),
\end{array}
\end{equation}
where $\bf{I}$ is an identity matrix, $s=p-1$, and $\Gamma(a, b)$ 
denotes a gamma distribution with shape parameter $a$ and rate parameter $b$, and $\tau_i^2$ and $\sigma_i^{-2}$
are independent. The normal-gamma-gamma (NGG) prior of model \eqref{EqBSEM} regularizes the parameter
estimates (e.g. operationalized as the posterior mean) in two distinct ways. First, due to the
normal prior on the regression coefficients $\beta_{i,r}$  (corresponding to a ridge penalty), the estimates of these parameters are
shrunken \textit{locally} (i.e.\ within each equation) to zero. Second, the estimates are
simultaneously shrunken \textit{globally} (i.e. across equations), due to the fact that  the hyperparameters
$\alpha =\{a_1,b_1,a_2,b_2\}$ do not depend on the index $i$. Here we have found a vague
prior on the error variances (e.g.\ $a_2=b_2=0.001$) to be appropriate to set the general scale of
the problem, whereas estimating the parameters $a_1,b_1$ in Empirical Bayes (EB) fashion
is advantageous, as it further ``borrows information'' across the regression equations. 
The resulting global shrinkage improves inference in particular for large networks
(see also Section~\ref{SectionNumericalInvestigation}). 
The BSEM model can be fit  computationally efficiently by a variational method,
and generally outperforms the aforementioned lasso
regression approach to the estimation of model \eqref{EqSEM}. Furthermore, variables can be accurately selected
based on the marginal posterior distributions of the regression coefficients (Leday et al., 2015).

The problem of network reconstruction is challenging due to the vast space of possible graphs for
even a moderate number of variables. This endeavour is further complicated by the inherent noise in
the measurements used for the reconstruction. Fortunately, network reconstruction need not
start from scratch, as often similar networks have been studied previously. Prior information on
the network may be available in the literature, repositories, or simply as pilot data.  It is natural to
take such information along in network reconstruction. This is already done in areas of (dynamic)
Bayesian networks. Among these studies, Werhli et al. proposed a framework to incorporate multiple sources of prior knowledge into dynamic Bayesian network using MCMC sampling (Werhli and Husmeier, 2007). In Bayesian network learning Murkherjee and Speed (2008) proposed a method to incorporate network features including edges, classes of edges, degree distributions, and sparsity using MCMC sampling. Isci et al. (2013) proposed also a framework to incorporate multiple sources of external knowledge in Bayesian network learning where the incorporation of external knowledge uses Bayesian network infrastructure itself. However,
none of these proposed methods accounts for the relevance of the prior knowledge.

In this paper we develop a method for incorporating external data or prior information into
the reconstruction of a conditional independence network. 
To this aim we extend in Section~\ref{SectionModel} the Bayesian SEM
framework \eqref{EqSEM}-\eqref{EqBSEM}. The extension incorporates prior knowledge in a flexible
manner. Next in Section~\ref{SectionVariationalBayes} we develop a variational Bayes approach to approximate the posterior
distributions of the regression parameters for given hyperparameters, and show this to be comparable in accuracy
to Gibbs sampling, although computationally much more efficient. In Section~\ref{SectionGlobalEB} this is complemented
by a derivation of an empirical Bayes approach to estimate the hyperparameters. 
Using simulations we show in Section~\ref{SectionNumericalInvestigation} that the method performs
better than competing methods that do not incorporate prior information (including BSEM)
when the prior knowledge is relevant, and is as accurate when it is not. In Section~\ref{SectionIllustration}	
we show the full potential of our approach on real data. We conclude the paper with a discussion.

\section{Model}
\label{SectionModel}
The BSEM approach, comprising model \eqref{EqSEM} with priors \eqref{EqBSEM}, is modified to incorporate external
information on the to-be-reconstructed network. The resulting model is referred to as BSEMed (BSEM
with \textit{e}xternal \textit{d}ata).

Prior knowledge on the network is assumed to be available as a ``prior network'', which specifies
which edges (conditional independencies) are present and absent. This is coded in an adjacency
matrix \text{P}, which contains only zeros and ones corresponding to the absence and presence of an
edge in the prior network.  That is, $\text{P}_{i,r}=1$ if node $i$ is connected with node $r$ and
$\text{P}_{i,r}=0$ otherwise. Note that the adjacency matrix P is symmetric (for the purpose of
undirected network reconstruction).

The BSEMed approach keeps equation \eqref{EqSEM}, but replaces the priors \eqref{EqBSEM} of BSEM by:
\begin{equation}
\label{EqBSEMed}
\begin{array}{ccc}
 \epsilon_i\given \sigma^2_i ,\tau_{i,0}^{2},\tau_{i,1}^{2}\sim \text{N}(0_n, \sigma_i^2{\bf{I}}_n),\\
 \beta_{i}\given \sigma_i^2,\tau_{i,0}^{2},\tau_{i,1}^{2} \sim \text{N}(0_s,\sigma_i^2{\bf{D}}_{\tau_i^{-2}}), \\
 {\bf{D}}_{\tau_i^{-2}}=\diag(\tau_{i,1}^{-2},...,\tau_{i,s}^{-2}) ,\\
 \tau_{i,r}^2=\begin{cases}
\tau_{i,0}^2 \sim \Gamma(a_0, b_0), & \mbox{if} \quad \text{P}_{i,r}=0,\\ 
\tau_{i,1}^2 \sim \Gamma(a_1, b_1), & \mbox{if} \quad \text{P}_{i,r}=1, 
\end{cases}\\
 \sigma^{-2}_i \sim \Gamma(a_2, b_2).
\end{array}
\end{equation}

The normal-gamma-gamma-gamma (NGGG) prior \eqref{EqBSEMed} retains the ridge-type regularization 
of the regression parameters $\beta_{i,r}$ of \eqref{EqBSEM}, through Gaussian priors on these
coefficients. The  crucial difference between the two priors reveals itself in the 
variances of the latter priors. For each regression equation $i$  there are two possible variances:
$$\beta_{i,r}\sim \begin{cases}
\text{N}(0,\sigma_i^2\tau_{i,0}^{-2}), & \mbox{if} \quad \text{P}_{i,r}=0,\\ 
\text{N}(0,\sigma_i^2\tau_{i,1}^{-2}), & \mbox{if} \quad \text{P}_{i,r}=1. 
\end{cases}
$$ 
Hence, the regression coefficients corresponding to edges that are present according to the prior
information share the same variance, and similarly for the other set of regression coefficients. Both variances
can be both small and large, as they are themselves modelled through Gamma priors, where small values lead
to small regression coefficients. If the prior information on the network were correct, then naturally
a small value of $\tau_{i,0}^{-2}$ would be desirable, smaller than the value of $\tau_{i,1}^{-2}$. However,
the construction is open-minded in that the two values, and even their priors, are not fixed
a-priori. In \eqref{EqBSEMed} the two parameters $\tau_{i,0}^{-2}$ and $\tau_{i,1}^{-2}$ receive
gamma priors, with different hyperparameters $(a_0,b_0)$ and $(a_1,b_1)$. For further flexibility
these hyperparameters will be adapted to the data by an empirical Bayes method. Then, if the absence of an edge in the prior network
is corroborated by the current data, the corresponding regression coefficient $\beta_{i,r}$
may stem from a prior with a small variance, and will tend to be small; a similar, but opposite, situation
will occur for edges that are present in the prior network. 
Indeed in Section~\ref{SectionNumericalInvestigation} we shall see that the EB approach will tend to find similar
values of $\tau_{i,0}^{2}$ and $\tau_{i,1}^{2}$ when the prior knowledge is non-informative,  and rather different values
otherwise.

\begin{figure}[!ht]
	\subfigure[Empty prior graph]{\includegraphics[width=0.4\textwidth]{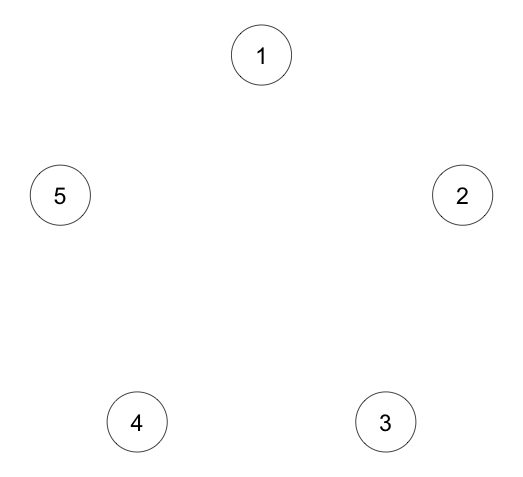}} 
	\subfigure[Complete prior graph]{\includegraphics[width=0.4\textwidth]{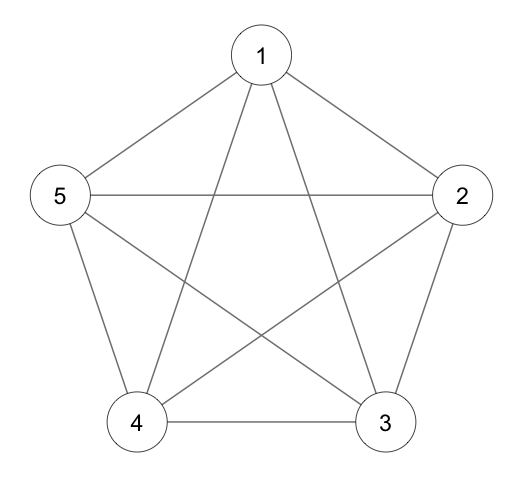}}
	\caption{Examples of two complementary graphs with $p=5$} 
	\label{FigureCompleteGraph}
\end{figure}

The fact that model \eqref{EqBSEMed} contains the model \eqref{EqBSEM} as a submodel,
harnesses against the misspecification of the prior information.
Although the number of latent variables in \eqref{EqBSEMed} is considerably higher (namely $p-1$ additional variances,
one for each regression equation),
the actual number of extra parameters is only two (the pair $(a_1,b_1)$). This explains that if the prior information is
incorrect or irrelevant for the data at hand, then the cost in terms of precision of the estimators is minor.
It is amply compensated by the gains if the prior information is correct. We corroborate this in our simulation study
in Section~\ref{SectionNumericalInvestigation}.
In this connection it is also of interest to note the interchangeable roles of $\tau^{2}_{i,0}$ and $\tau^{2}_{i,1}$,
which causes pairs of complementary prior networks to lead to exactly the same
``posterior network''. For instance,  the empty and complete graphs (see Figure~\ref{FigureCompleteGraph})
boil down to the same prior.

\section{Variational Bayes method and Gibbs sampling}
\label{SectionVariationalBayes}
In this section we develop a variational Bayes approach to approximate the (marginal) posterior distributions 
of the parameters $\beta_{i,r}, \tau_{i,0}^2, \tau_{i,1}^2, \sigma_i^2$ in model \eqref{EqBSEMed}. The algorithm
is similar, but still significantly different, from the algorithm developed in Leday et al. (2015) for the model \eqref{EqBSEM}. In the following we can see that, due to \eqref{EqBSEMed}, the variational parameters have a form which renders the implementation of \eqref{EqBSEMed} much more challenging.
We also verify that these approximations are accurate by comparing them to the results obtained using a Gibbs sampling strategy, which
is much slower. Computational efficiency is an important characteristic, especially for fitting large networks.

In this section we work on a single regression equation, i.e.\ for a fixed index $i$, and given hyperparameters $a_k, b_k$, for $k=0,1,2$.
In the next section we combine the regression equations to estimate the hyperparameters.

\subsection{Variational Bayes inference.} 
In general a ``variational approximation'' to a distribution is simply the closest element in a given
target set $\mathcal{Q}$ of distributions, usually with ``distance'' measured by Kullback-Leibler divergence.
The set $\mathcal{Q}$ is chosen both for its computational tractability and accuracy of approximation. Distributions $Q$ with
stochastically independent marginals (i.e.\ product laws) are popular, and then the ``accuracy'' of approximation \
is naturally restricted to the marginal distributions.

In our situation we wish to approximate the posterior distribution of the parameter
$\theta:=(\beta_i,\tau_{i,0}^2, \tau_{i,1}^2, \sigma_i^2)$ given the prior \eqref{EqBSEMed} and the observation $Y_i$ given in \eqref{EqSEM},
for a fixed $i$. Here in \eqref{EqSEM} we take $X_i$ (which depends on $Y_j$ for $j\not=i$) as given, 
as in a fixed-effects linear regression model. For $p(\cdot\given Y_i)$ the posterior density in this model,
the variational Bayes approximation is given as 
$$q^* = \argmin_{q\in \mathcal{Q}} \mathbf{E}_q \log \frac{q(\theta)}{p(\theta\given Y_i)},$$
where the expectation is taken with respect to the density $q\in \mathcal{Q}$.
For $p(Y_i,\theta)$ the joint density of $(Y_i,\theta)$, this is equivalent to finding the maximizer of 
\begin{equation}
\label{EqLowerBound}
 \mathbf{E}_q\log \frac{p(Y_i,\theta)}{q(\theta)}.
\end{equation}
By the nonnegativity of the Kullback-Leibler divergence, the latter expression is a lower bound 
on the marginal density $p(Y_i)=\int p(Y_i,\theta)\,d\theta$ of the observation,
and consequently is usually referred to as ``the lower bound''. Solving the variational problem
is equivalent to maximizing this lower bound (over $\mathcal{Q}$).

We choose the collection $\mathcal{Q}$ equal to the set of distributions of
$\theta$ for which the components $\beta_i$, $\tau_{i,0}^2$, $\tau_{i,1}^2$ and $\sigma_i^2$ are stochastically independent,
i.e.\ $q(\theta) = \prod_{l=1}^4 q_l(\theta_l)$, where the marginal densities $q_l$ are arbitrary. Given such a factorization of $q$
it can be shown in general (see e.g. Ormerod and Wand (2010)), that the optimal marginal densities $q^*_l$ satisfy:
$$q^*_l(\theta_l) \propto \exp(\mathbf{E}_{q_{\textbackslash l}}\log p(Y_i, \theta)),
\qquad\text{ where }\mathbf{E}_{q_{\textbackslash l}} = \mathbf{E}_{q_1}\ldots \mathbf{E}_{q_{l-1}}\mathbf{E}_{q_{l+1}}\ldots \mathbf{E}_{q_4}.$$
It can be shown (see the Supplementary Material)
that in model \eqref{EqBSEMed} for regression equation $i$, with $\theta = (\beta_i, \tau^2_{i,0},\tau^2_{i,1}, \sigma^{-2}_i)$,
this identity can be written in the concrete, ``conjugate'' form
\begin{equation}
\begin{aligned}
 \beta_i\given Y_i &\sim \text{N}\big(\beta_i^* ,\Sigma_i^*\big),\\
\tau^2_{i,0}\given Y_i& \sim \Gamma\big(a_{i,0}^*, b_{i,0}^*\big),\\
\tau^2_{i,1}\given Y_i &\sim \Gamma\big(a_{i,1}^*, b_{i,1}^*\big),\\
\sigma^{-2}_i\given Y_i& \sim \Gamma\big(a_{i,2}^*, b_{i,2}^*\big),
\end{aligned} 
\label{EqVarBayesIterations}
\end{equation}
where 
\begin{align*}
 \Sigma_i^\ast &= \big[\mathbf{E}_{q_4^*}(\sigma_i^{-2})\big(X_i^TX_i + {\bf{D}}_{\mathbf{E}_{q^\ast_2\cdot q^\ast_3}(\tau_i^2)}\big)\big]^{-1}, \\
\beta_i^* &= \big[X_i^TX_i + {\bf{D}}_{\mathbf{E}_{q^\ast_2\cdot q^\ast_3}(\tau_i^2)}\big]^{-1}X_i^TY_i,
\end{align*}
\begin{align*}
a_{i,0}^*&=a_0+\thalf{s^0}, \qquad &b_{i,0}^*&=b_0+\thalf\mathbf{E}_{q^\ast_4}(\sigma_i^{-2})\mathbf{E}_{q^\ast_1}({\beta_i^0}^T\beta_i^0),\\
a_{i,1}^*&=a_1+\thalf{s^1}, \qquad &b_{i,1}^*&=b_1+\thalf\mathbf{E}_{q^\ast_4}(\sigma_i^{-2})\mathbf{E}_{q^\ast_1}({\beta_i^1}^T\beta_i^1),\\
a_{i,2}^*&=a_2+\thalf{n}+\thalf{s},\quad  &b_{i,2}^*&=b_2 +\thalf\mathbf{E}_{q^\ast_{\textbackslash 4}}\big(\beta_i^T{\bf{D}}_{\tau_i^2}\beta_i\big)\\
&&&\qquad\qquad+ \thalf \mathbf{E}_{q^\ast_1}(Y_i- X_i\beta_i)^T(Y_i-X_i\beta_i),
\end{align*}
where $s^0$ and $s^1$ are the number of $0$'s and $1$'s in the $i$-th row of the adjacency matrix
\text{P}, not counting the diagonal element; and
$\beta_i^0 =\{\beta_{i,r}: r\in \mathcal{I}\backslash i, \text{P}_{i,r}=0 \}$ 
and $\beta_i^1 =\{\beta_{i,r}: r\in \mathcal{I}\backslash i, \text{P}_{i,r}=1 \}$ are the coordinates of the vector of regression parameters
corresponding to these $0$'s and $1$'s.
Furthermore 
$${\bf{D}}_{\mathbf{E}_{q^\ast_2\cdot q^\ast_3}(\tau_i^2)} 
= \diag\Bigl(\mathbf{E}_{q^\ast_2} \mathbf{E}_{q^\ast_3}  (\tau_{i,1}^2),...,\mathbf{E}_{q^\ast_2}\mathbf{E}_{q^\ast_3}  (\tau_{i,s}^2)\Bigr).$$
In these identities the optimal densities $q_l^*$ appear both on the left and the right of the equations and hence
the identities describe the optimal densities only as a fixed point. In practice the identities are iterated ``until convergence''
from suitable starting values. 

The iterations also depend on the hyperparameters $a_k, b_k$. In the next section
we describe how these parameters can be estimated from the data by blending in updates of these parameters in
the iterations.

\subsection{Variational Bayes vs Gibbs sampling.}
Under the true posterior distribution the coordinates $\beta_i,\tau_{i,0}^2, \tau_{i,1}^2, \sigma_i^2$ are not independent.
This raises the question how close the variational approximation is to the true posterior distribution.
As the latter is not available in closed form, we investigate this question in this section by comparing
the variational approximation to the distribution obtained by running a Gibbs sampling algorithm for a long time. 
As for the network reconstruction we only use the marginal posterior distributions of the regression parameters,
we restrict ourselves to these marginal distributions.

The full conditional densities of BSEMed can be seen to take the  explicit form:
\begin{align*}
 \beta_i\given Y_i,\tau_{i,0}^{2},\tau_{i,1}^{2},\sigma_i^{-2} &   \sim \text{N}(\beta_i^* ,\Sigma_i^*), \\
\tau_{i,0}^{2}\given Y_i,\beta_i,\tau_{i,1}^{2},\sigma_i^{-2} &                                  \sim \Gamma(a_{i,0}^*, b_{i,0}^*), \\
\tau_{i,1}^{2}\given Y_i,\beta_i,\tau_{i,0}^{2},\sigma_i^{-2} &                                  \sim \Gamma(a_{i,1}^*, b_{i,1}^*), \\
\sigma_i^{-2}\given Y_i,\beta_i,\tau_{i,0}^{2},\tau_{i,1}^{2} &                            \sim \Gamma(a_{i,2}^*, b_{i,2}^*),
\end{align*}

\begin{figure}
	\subfigure[Betas2to9, Tau0, Tau1, Sigma]{\includegraphics[width=0.9\textwidth]{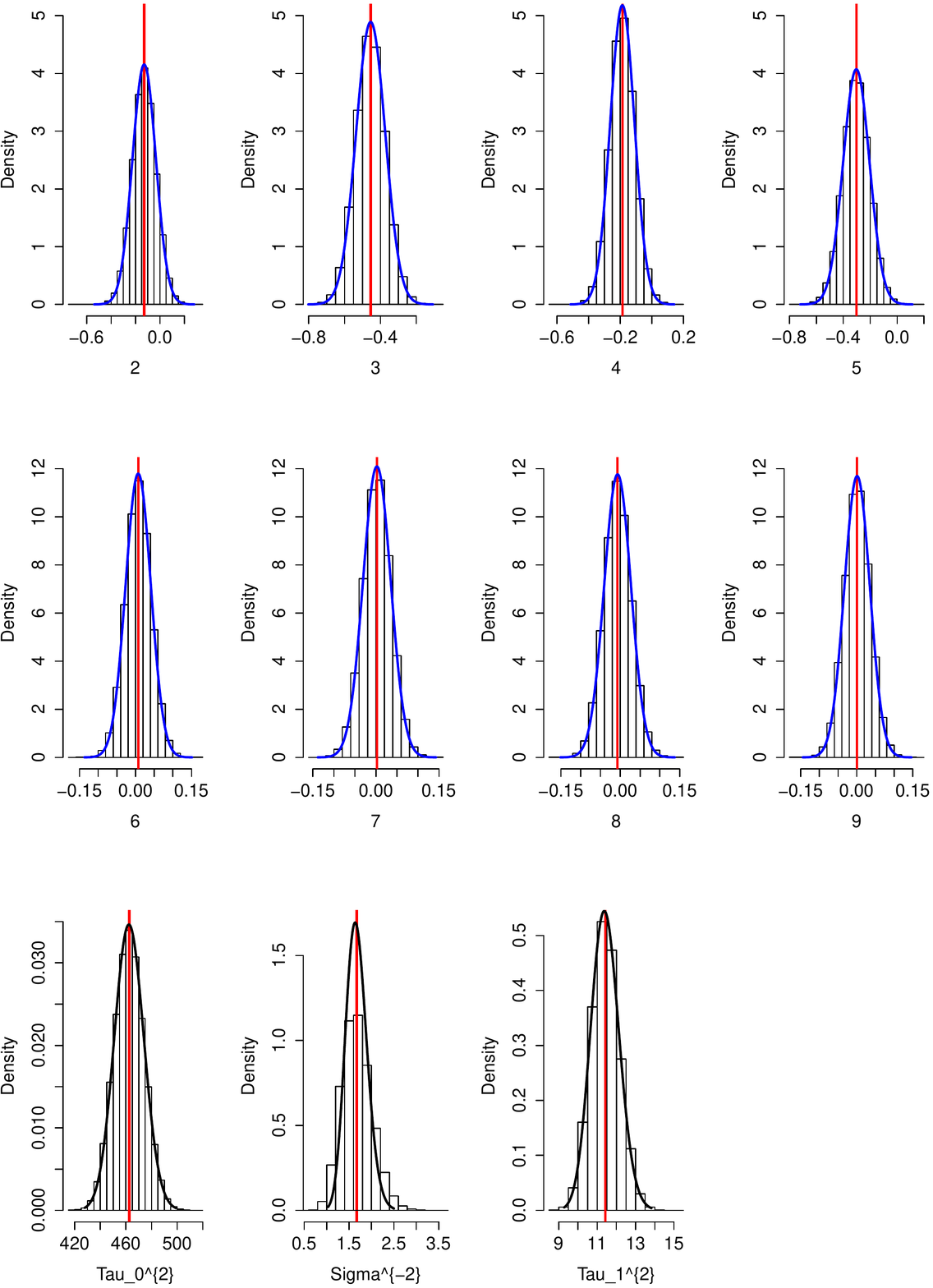}} 
	\caption{Comparison of variational marginal densities of $\beta_{1,2},\ldots,\beta_{1,9}$ (blue curves) with corresponding Gibbs sampling-based histograms on the one hand and variational marginal densities of $\tau_{1,0}^2$, $\tau_{1,1}^2$ and $\sigma_1^{-2}$ (black curves) also with corresponding Gibbs sampling-based histograms on the other hand. The red vertical lines display the variational marginal means.} 
	\label{FigureGibbs}
\end{figure}

where the parameters $\Sigma_i^*$, $\beta_i^*$, $a_{i,k}^*$ and $b_{i,k}^*$ satisfy the same system of equations
as in the variational algorithm, except that all expectations $\mathbb{E}_{q^\ast}$ must be replaced
by the ``current'' values  taken from the conditioning (see Supplementary Material). Thus Gibbs sampling of the full posterior
$(\beta_i,\tau^{2}_{i,0},\tau^{2}_{i,1}, \sigma^{-2}_i)\given Y_i$ is easy to implement, although slow.

We ran a simulation study with a single regression equation (say $i=1$) with $n=p=50$,
and compared the variational Bayes estimates of the marginal densities with the corresponding Gibbs sampling-based
estimates. Thus we sampled $n=50$ independent replicates from a $p=50$-dimensional normal distribution with mean zero
and $(p\times p)$-precision matrix $\Omega$, and formed the vector $Y_1$ and matrix $X_1$ as indicated in \eqref{EqSEM}.
The precision matrix was chosen to be a \emph{band matrix} with a lower bandwidth $b_l$ equal to the upper bandwith $b_u$. It is $b_l=b_u=4$, thus a total number of $9$ band elements including the diagonal.
For both the variational approximation and the Gibbs sampler we used prior hyperparameters $a_2=b_2=0.001$ 
and prior hyperparameters $\hat{a}_0, \hat{b}_0, \hat{a}_1, \hat{b}_1$ fixed to the values
set by the \emph{global}  empirical Bayes method described in Section~\ref{SectionGlobalEB}.
The Gibbs iterations were run $nIter=100,000$ times, after which the first $nBurnin=1000$ iterates were discarded. 
In Figure~\ref{FigureGibbs} we plot histograms based on subsampling every 10th value of the iterations,
with the variational Bayes approximation to the marginal posterior densities overlaid as a curve. To save space
we only plot the densities of $\beta_{1,2},\ldots,\beta_{1,9}$ and $\tau_{1,0}^2,\tau_{1,1}^2,\sigma_1^{-2}$; the
plots of the densities of $\beta_{1,10},\ldots,\beta_{1,50}$ are very similar. The correspondence between the two
methods is remarkably good.

We conclude that the variational Bayes method gives reliable estimates of the posterior marginal distributions.
Table~\ref{TableGibbs} compares the computing times for the two methods (in R). The variational method
clearly outperforms the Gibbs sampling method, which would hardly be feasible even for $n=p=50$.

\begin{table}
	\begin{center}  
		\begin{tabular}{|c | c | c | c|}
			\hline
			& BSEMed      &   Gibbs sampling   \\  \hline 
			time in seconds           &    40         &  $2542\times50= 127,100$           \\ 
			\hline
		\end{tabular}
		\captionof{table}{Computing times for an R-implementation of the variational Bayes method and the Gibbs sampling method
			applied to \eqref{EqSEM} with $n=p=50$.}
		\label{TableGibbs}
	\end{center}
\end{table}

\section{Global empirical Bayes for BSEMed}
\label{SectionGlobalEB}
Model \eqref{EqBSEMed} possesses three pairs of  hyperparameters $(a_k,b_k)$, for $k\in\{0,1,2\}$.
The pair $(a_2,b_2)$ controls the prior of the error variances $\sigma_i^2$; we fix this 
to numerical values that render a vague prior, e.g.\ to $(0.001, 0.001)$. In contrast, we let the values
of the parameters  $\alpha=(a_0,b_0,a_1,b_1)$ be determined by the data. As these hyperparameters
are the same in every regression model $i$, this allows information to be borrowed across the regression equations,
leading to  \emph{global shrinkage} of the regression parameters. 

A natural method to estimate the parameter $\alpha$ is to apply maximum likelihood to the marginal likelihood
of the observations in the Bayesian BSEMed model determined by \eqref{EqSEM} and \eqref{EqBSEMed}.
Here ``marginal'' means that all parameters except $\alpha$ are integrated out of the likelihood according to their prior. 
The approach is similar to the one in van de Wiel et al. (2012).
As a first simplification of this procedure we treat the vectors $Y_1,\ldots, Y_p$ as independent, thus
leading to a likelihood of product form. As the exact marginal likelihoods of the $Y_i$ are intractable, 
we make a second simplication and replace these likelihoods by the lower bound 
\eqref{EqLowerBound} to the variational Bayes criterion (see Supplementary Material). 

Recall that in model \eqref{EqBSEMed} each regression parameter $\beta_{i,r}$ corresponds to one of two normal priors, that is:

$$\beta_{i,r}\sim 
\begin{cases}
\text{N}(0,\sigma_i^2\tau_{i,0}^{-2}), & \mbox{if} \quad \text{P}_{i,r}=0,\\
\text{N}(0,\sigma_i^2\tau_{i,1}^{-2}), & \mbox{if} \quad \text{P}_{i,r}=1.
\end{cases}$$

It is the regression coefficients corresponding to edges that are not present according to the prior
information share the same precision $\tau_{i,0}^2$, and similarly the coefficients corresponding to the edges that are present obtain the precision $\tau_{i,1}^2$. Both precisions receive gamma priors with different hyperparameters that are adapted by the current data by the means of the global EB procedure described above. Then, if the absence of an edge in the prior network
is corroborated by the current data, the corresponding regression coefficient $\beta_{i,r}$
may stem from a prior with a small variance, and will tend to be small; a similar, but opposite, situation
will occur for edges that are present in the prior network. 
In next Section we shall see that the EB approach will tend to find similar
values of $\tau_{i,0}^{2}$ and $\tau_{i,1}^{2}$ when the prior knowledge is non-informative,  and rather different values
otherwise.

\section{Numerical investigation}
\label{SectionNumericalInvestigation}
To study the effect of including a prior network in the model framework we compare BSEMed with BSEM.
Hereto, we generated data $Y^1,\ldots, Y^n$ according to \eqref{EqData}, for $p=100$ and
$n \in \{50,200\} $, which reflect a high- and a low-dimensional situation, respectively. We
considered precision matrices $\Omega_p$, which imply {\em{band}}, {\em{cluster}} and {\em{hub}}
network topologies.

\begin{figure}[!]
	\subfigure[Band: n = 50 ]{\includegraphics[width=0.39\textwidth]{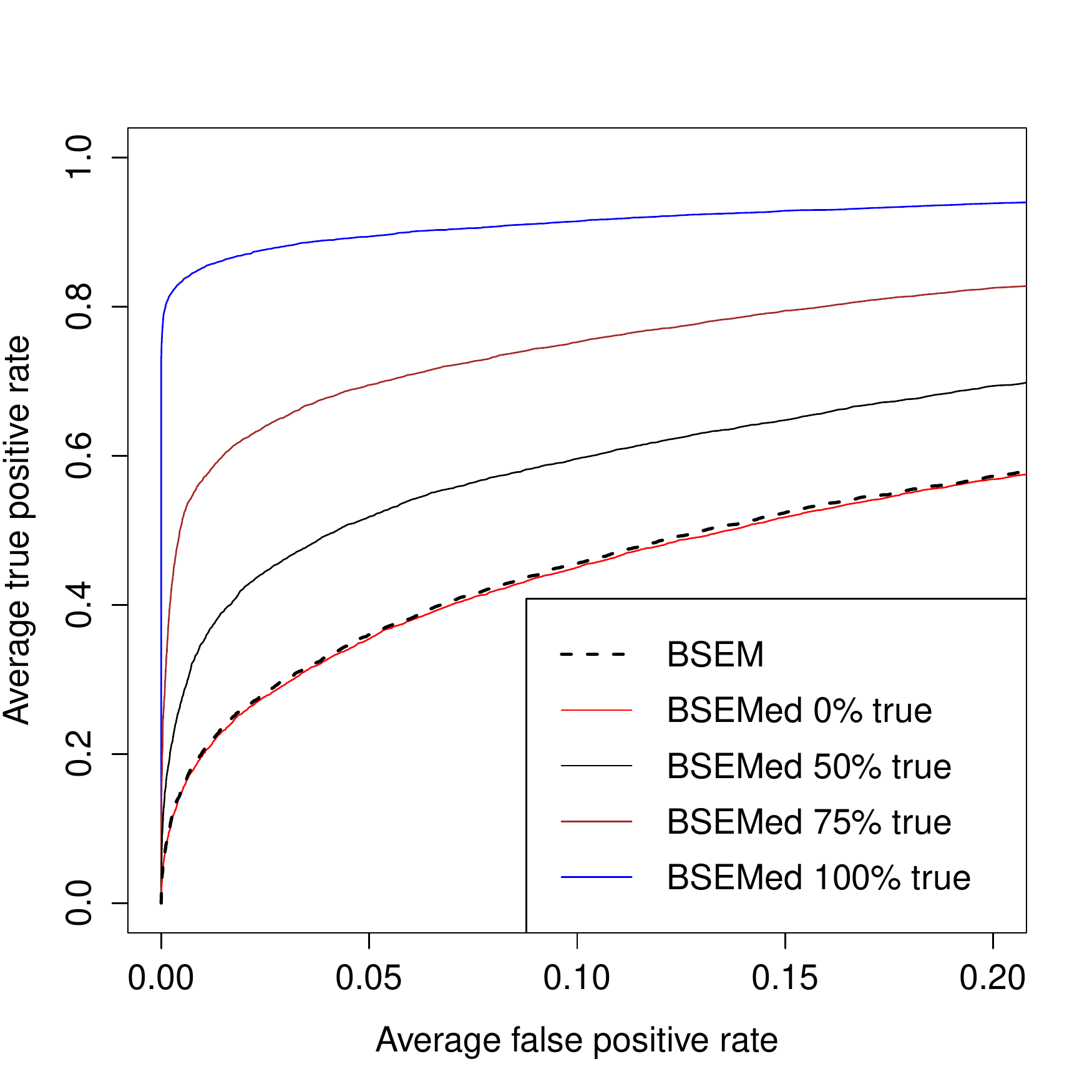}}
	\subfigure[Band: n = 200]{\includegraphics[width=0.39\textwidth]{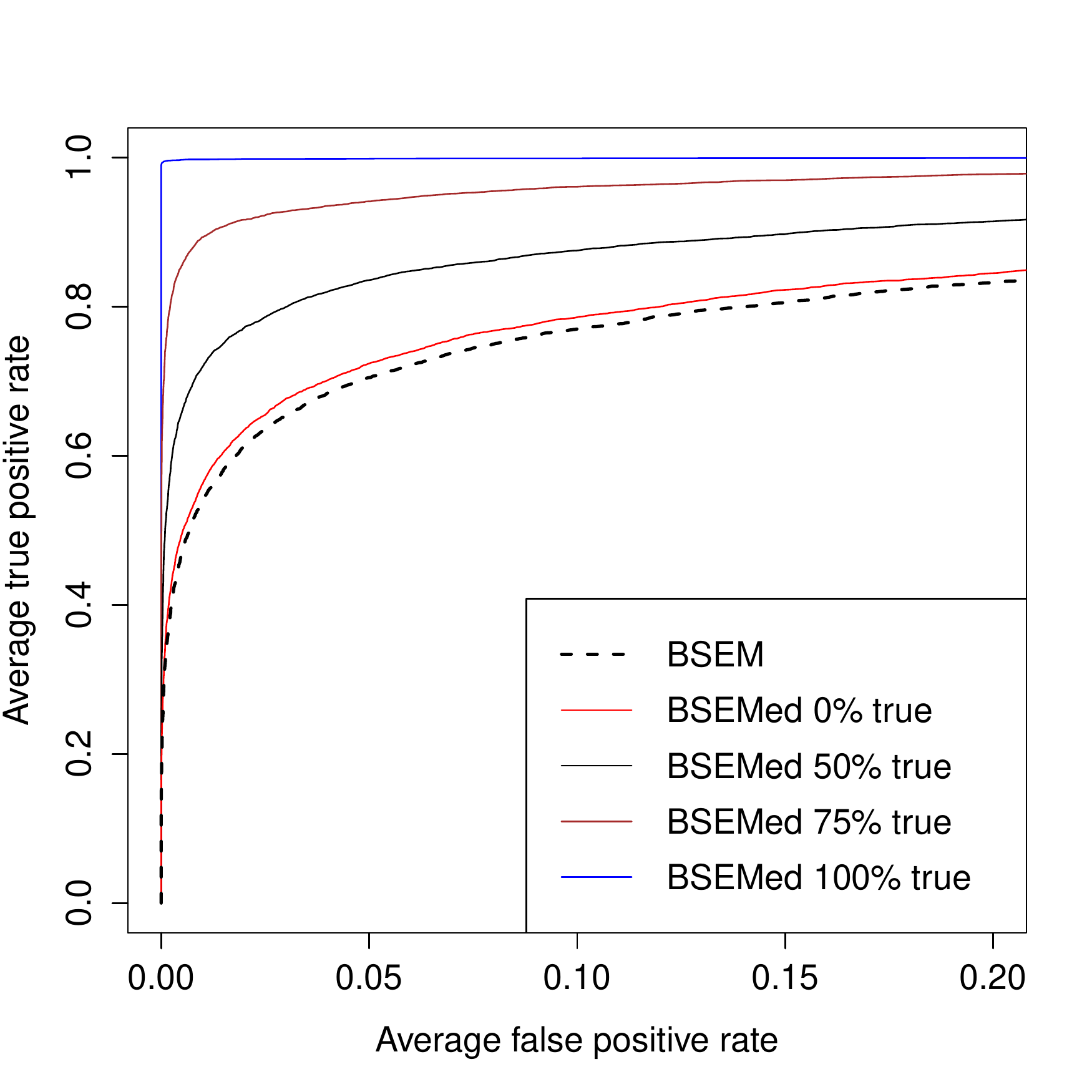}}
	
	\subfigure[Cluster: n =50]{\includegraphics[width=0.39\textwidth]{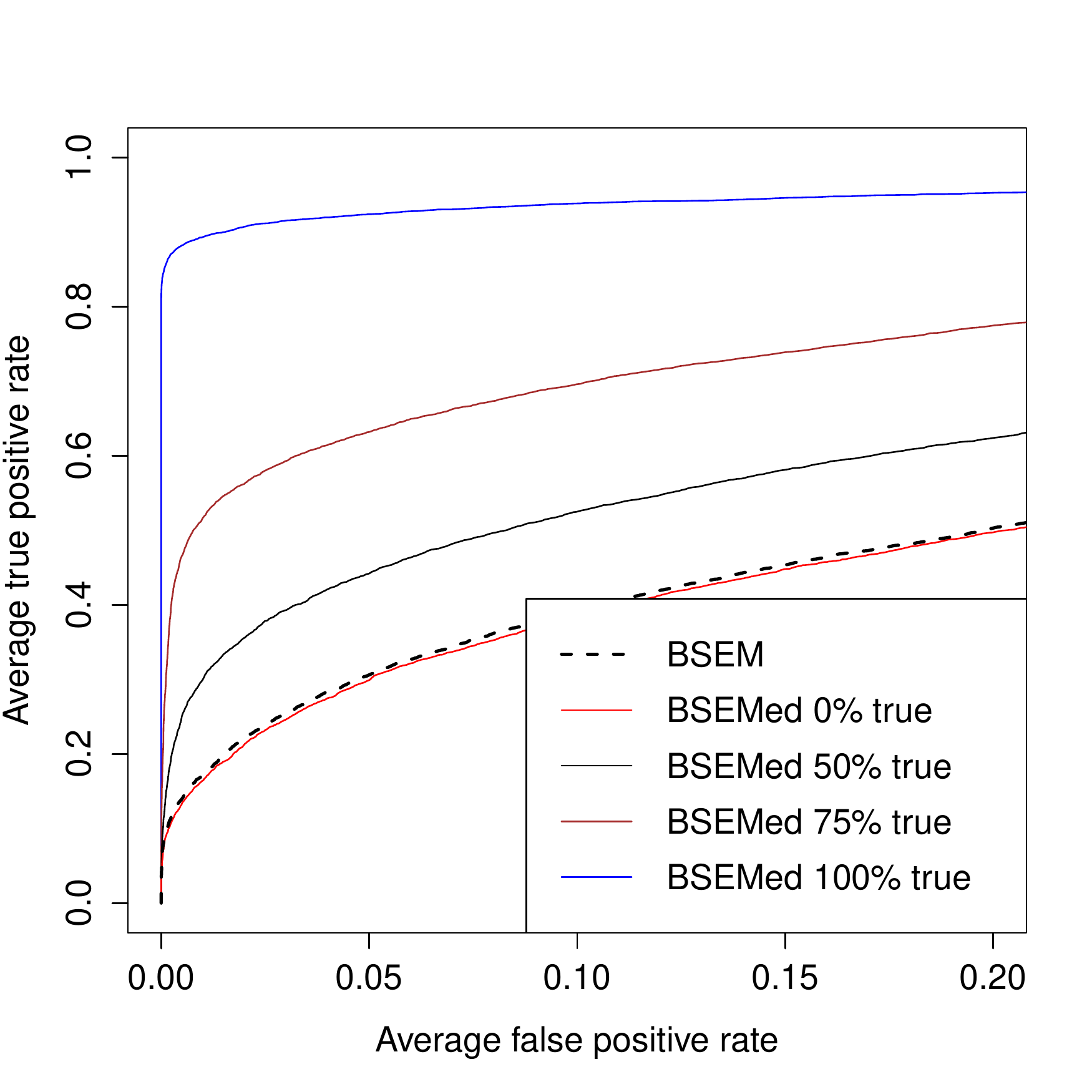}}
	\subfigure[Cluster: n = 200]{\includegraphics[width=0.39\textwidth]{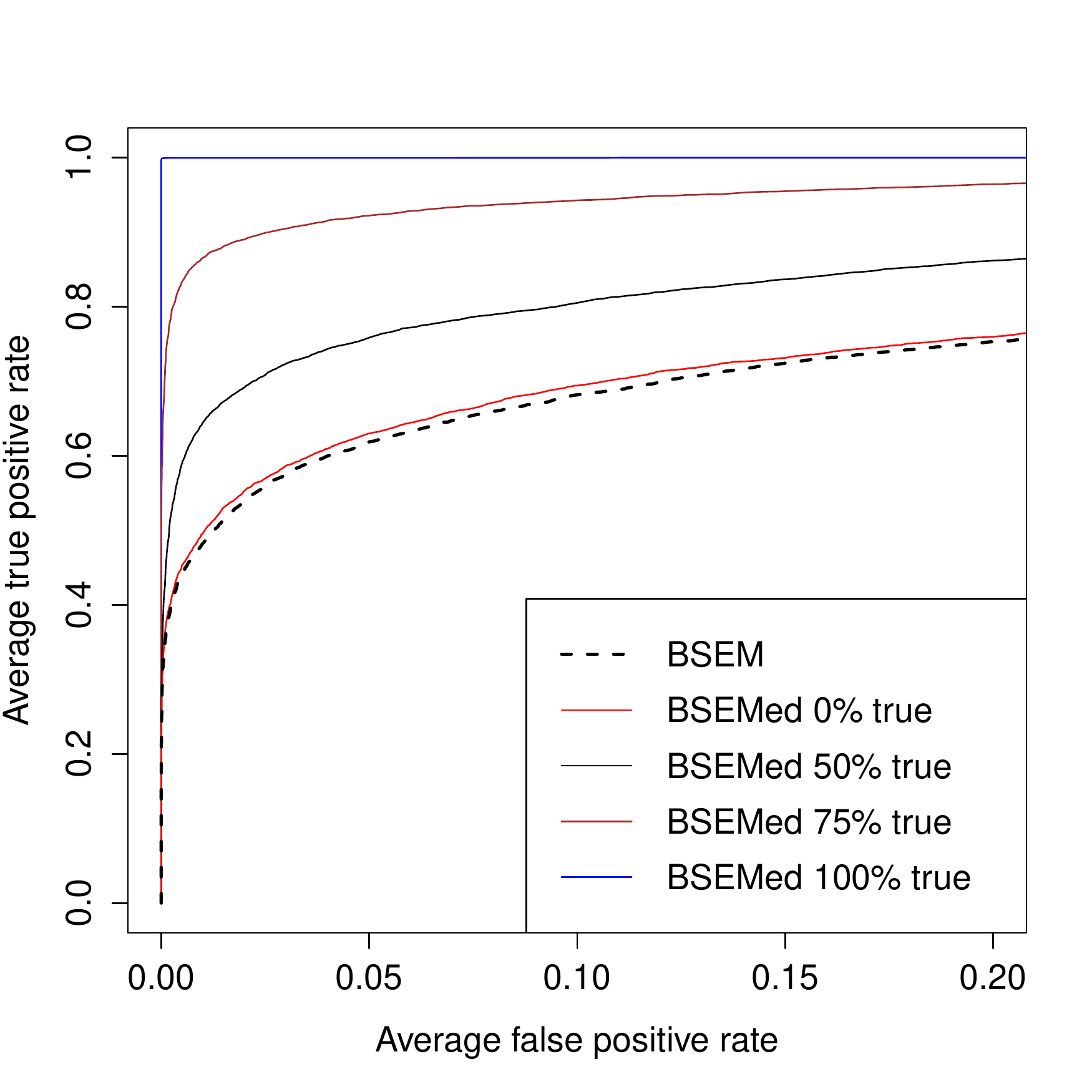}}
	
	\subfigure[Hub: n = 50]{\includegraphics[width=0.39\textwidth]{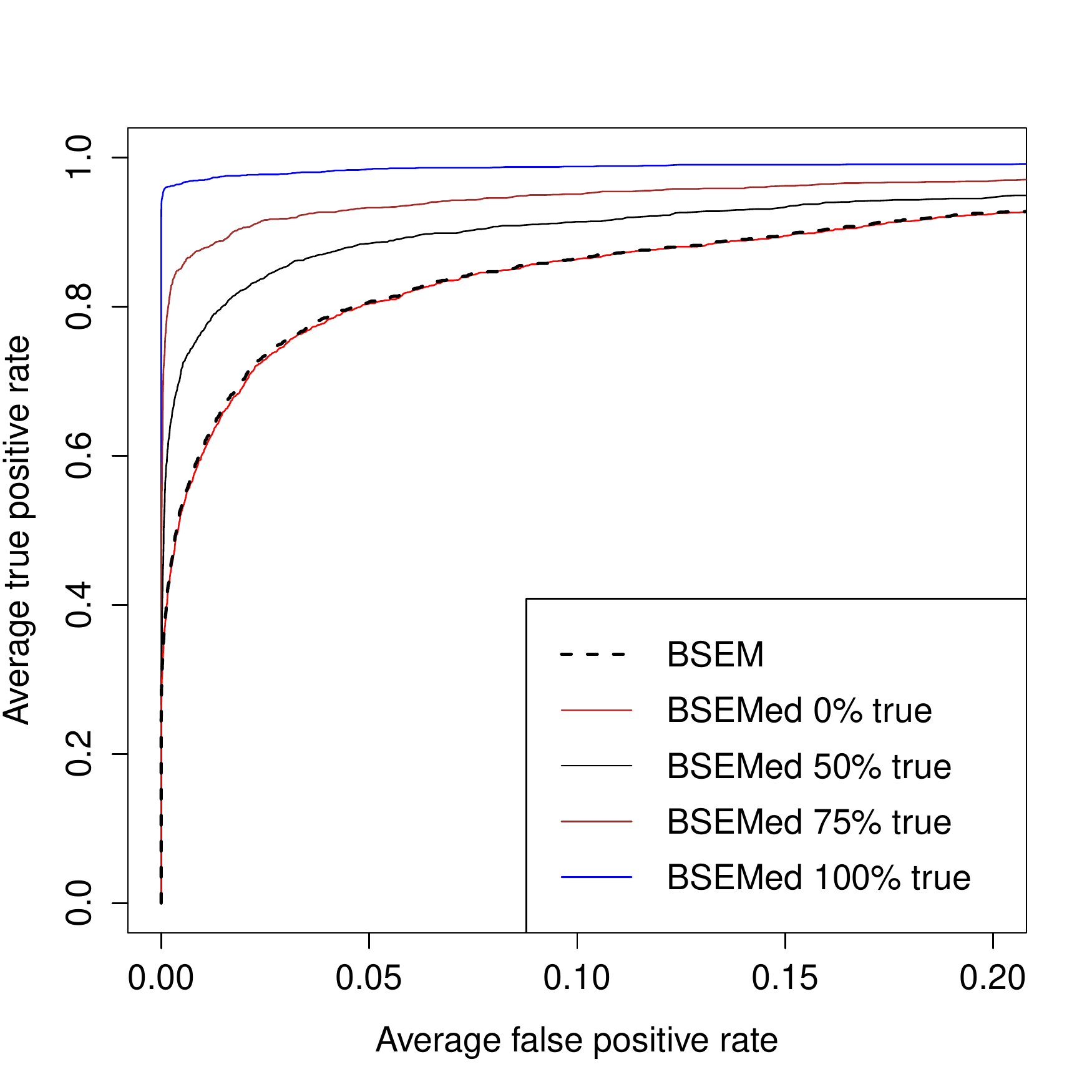}}
	\subfigure[Hub: n = 200]{\includegraphics[width=0.39\textwidth]{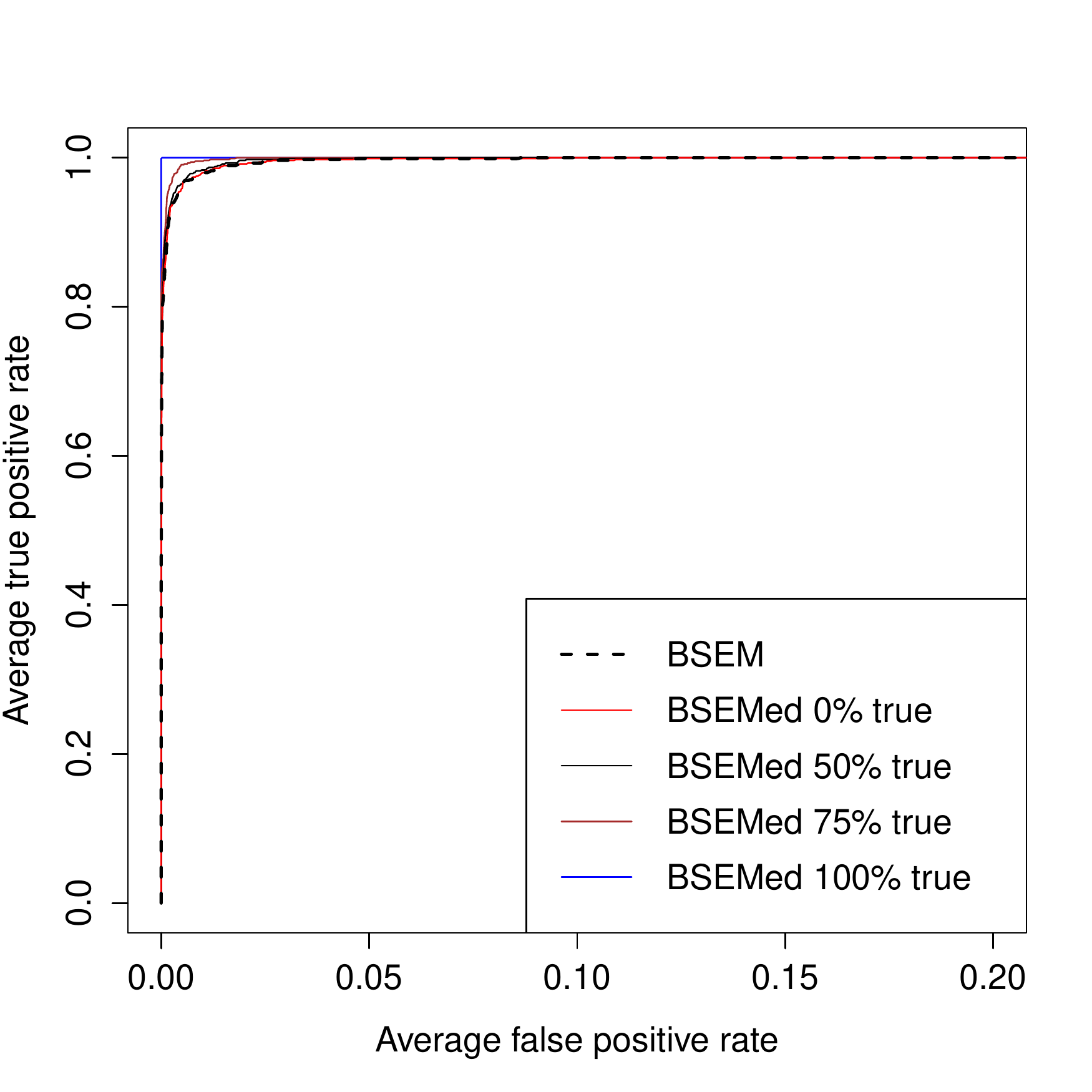}}
	\caption{ROC curves for BSEM (dashed) and BSEMed using perfect prior information (blue), BSEMed using 75\% true edges information (brown), BSEMed using 50\% true edges information (black) and BSEMed using 0\% true edges information (red). Here, $p=100$ and $n \in \{50,200\}$. }\label{rocs}
\end{figure}

For BSEMed we vary the quality of the prior network information: `perfect' prior information,
i.e. the generating model; `$75\%$' true edges; `$50\%$' true edges; `$0\%$' true edges.  To generate $75\%$  (or $50\%$, or $0\%$) true
information, we swapped $25\%$ (or $50\%$, or $100\%$) of the true edges with the same number of absent
edges, i.e.\ in the adjacency matrix $\text{P}$ that describes the prior network we swapped
these percentages of $1$s with $0$s. It may be noted that in the last case the prior network is completely wrong for the true edges,
but not for the absent edges due to over-sampling of the 0's, which seems realistic. Each simulation
is repeated 50 times. We display the performances of BSEM and BSEMed by ROC curves, which depict the
average false positive rate against the average true positive rate (see Figure~\ref{rocs}). We
observe from Figure~\ref{rocs} that BSEMed performs better than BSEM when the prior information is
relevant and as good as BSEM when the prior is wrong. The latter reflects the adaptive nature of the
EB procedure.

\begin{figure}
	\subfigure[True graph]{\includegraphics[width=0.48\textwidth]{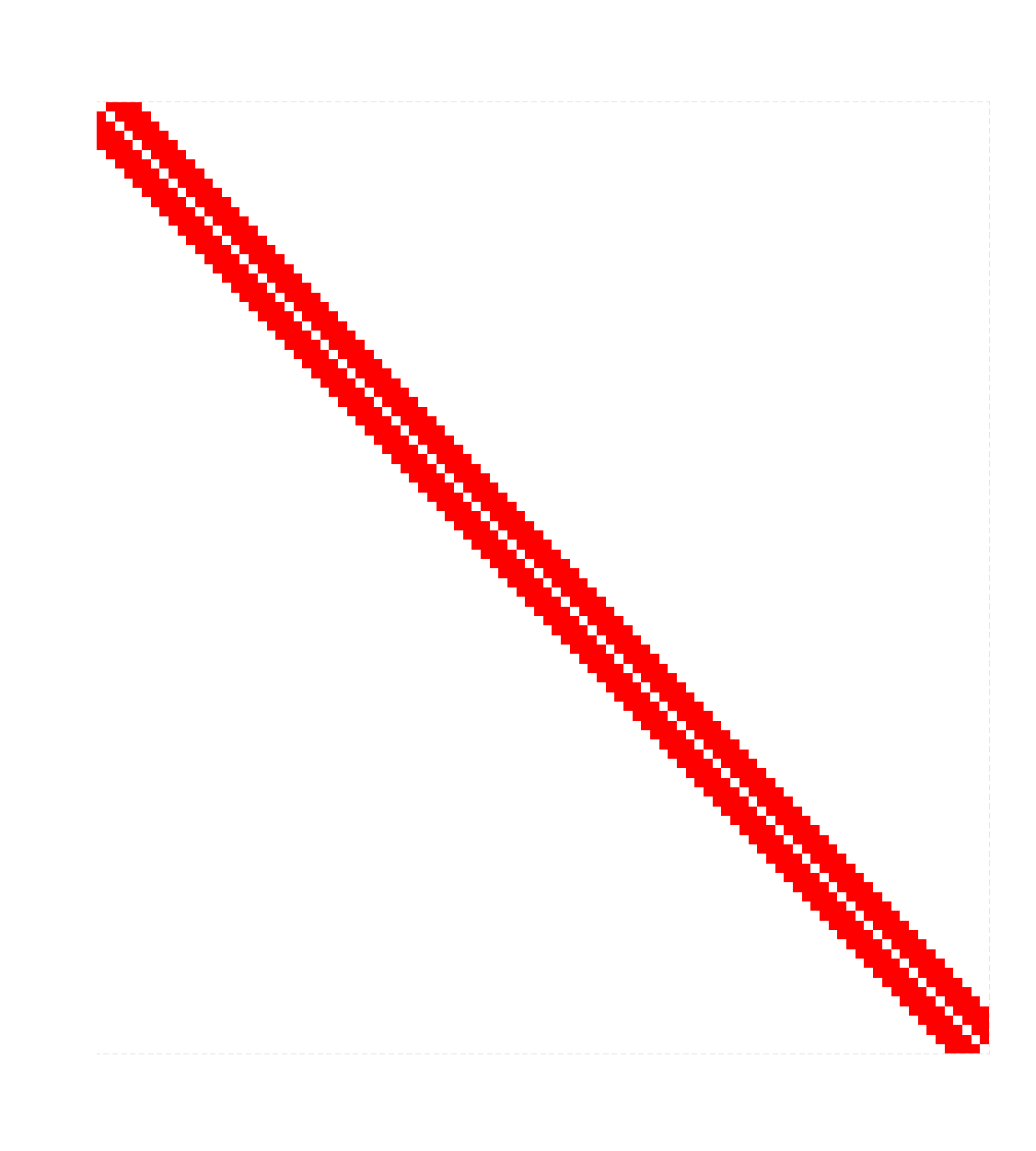}}
	\subfigure[BSEMed: perfect prior]{\includegraphics[width=0.48\textwidth]{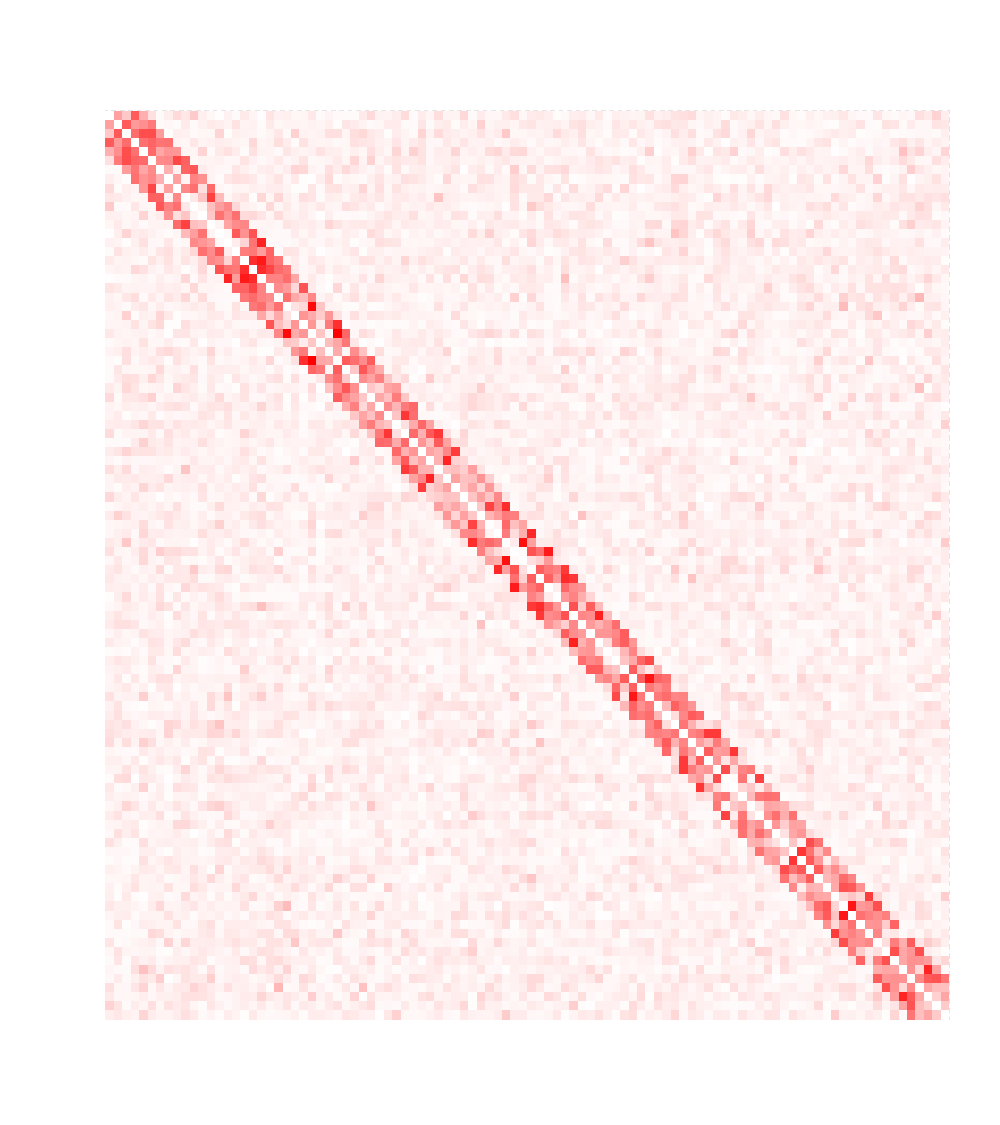}}
	\subfigure[BSEMed: 50 \% true Info]{\includegraphics[width=0.48\textwidth]{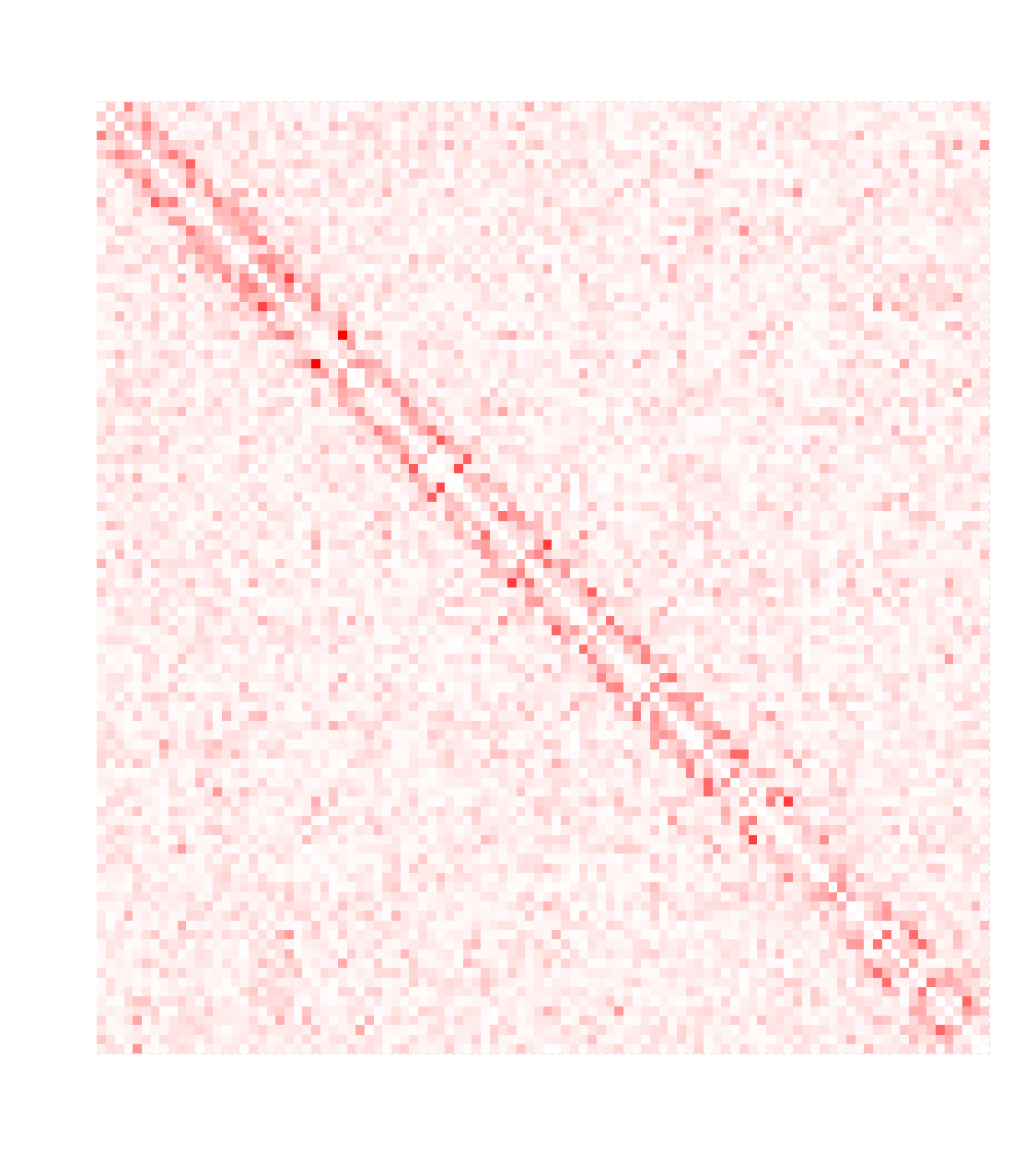}}
	\subfigure[BSEM]{\includegraphics[width=0.48\textwidth]{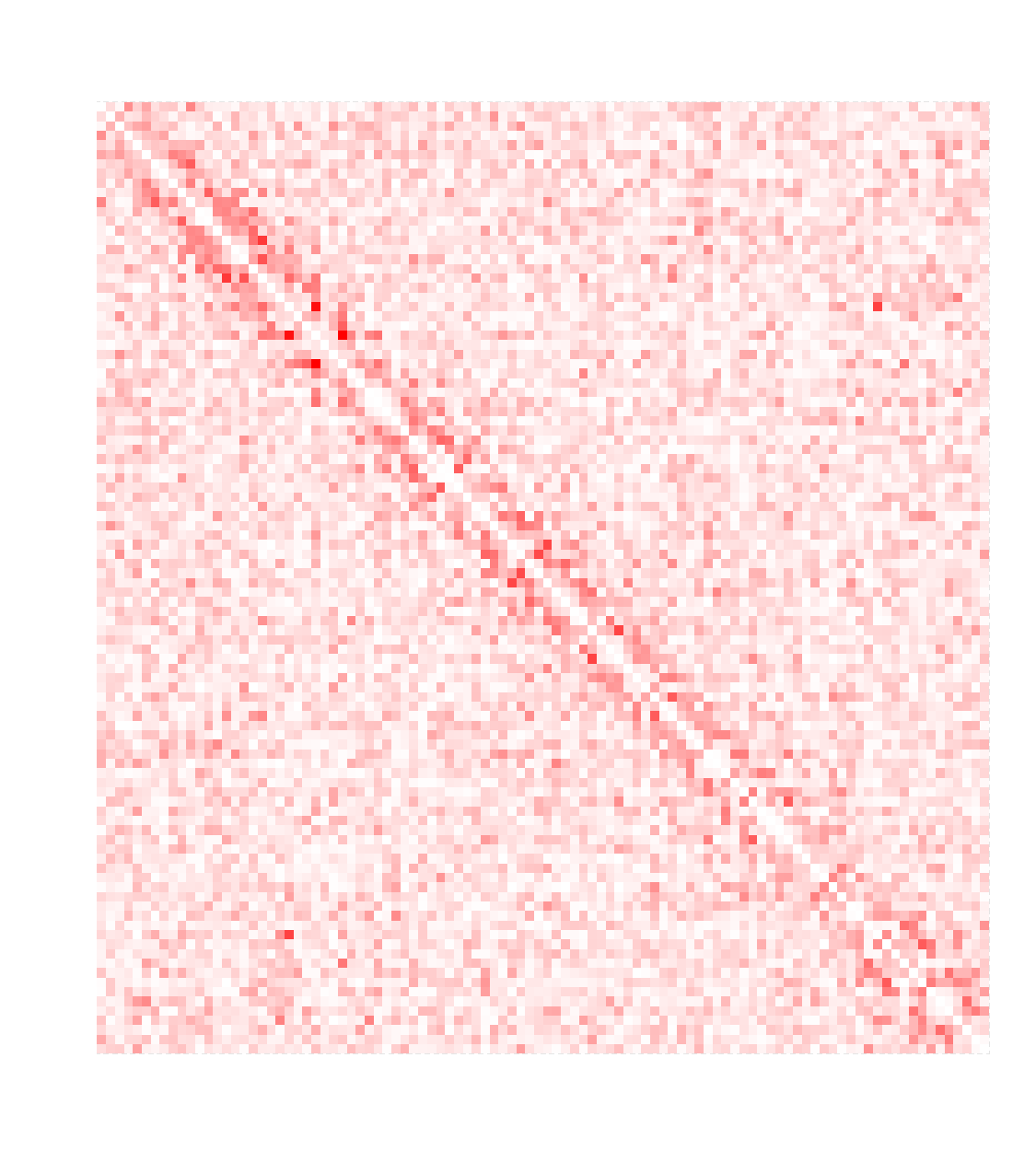}}
	\caption{Visualization of BSEMed estimate using perfect prior (b), BSEMed estimate using 50\% true edges information (c), BSEM estimate (d) and the true graph (a) in case $n=50$ and $p=100$.}\label{graphs}	
\end{figure}

We also consider the EB estimates. 
We summarize the precisions by their prior means, as estimated by the EB procedure:
$E(\tau^{2}_{i,k})={\hat{a}_k}/{\hat{b}_k}$, for $k\in\{0,1\}$.  When prior knowledge is relevant,
we expect ${\hat{a}_0}/{\hat{b}_0} > {\hat{a}_1}/{\hat{b}_1}$.  In the case with $0\%$ true edges,
the prior is irrelevant and partly wrong: none of the truly present edges are in the prior network
while some of the truly absent edges are part of the prior network. Hence, we expect the EB
procedure to render ${\hat{a}_1}/{\hat{b}_1}$ slightly larger than ${\hat{a}_0}/{\hat{b}_0}$. As
discussed in Section~\ref{SectionModel} for the complementary case, reversal of the roles of the two
priors can still improve performance of BSEMed, or at least not deteriorate it.

The EB estimates of the prior means are presented in Table \ref{EBestimates} for the case corresponding to Figure \ref{rocs}(a): {\em{band}} structure, $n=50$.

\begin{center}				
	\begin{tabular}{|c | c | c | c|}
		\hline
		& ${\hat{a}_0}/{\hat{b}_0}$ &  ${\hat{a}_1}/{\hat{b}_1}$  &       ratio \\  \hline
		true                  &   366.10       &      8.08    &   45.30        \\ \hline 
		$0.75\%$ true  edges &     272.97     &     14,36  &    19.00    \\  \hline
		$0.50\%$ true  edges &  216.10        &     27.56  &    7.84    \\  \hline
		$0\%$ true edges & 142.59         &    152.95  &      1.07     \\
		\hline
	\end{tabular}
	\captionof{table}{EB estimates of the prior means of precisions $\tau_{i,0}^2$ and $\tau_{i,1}^2$ in case of the {\em{band}} structure and $n=50$ for various qualities of prior information}\label{EBestimates}
\end{center}
Table \ref{EBestimates}  displays the prior means of precision, as estimated by EB, for BSEMed models with various qualities of prior information. It is clear that the better the quality of the prior information is, the larger the ratio of mean prior precisions is.

Figure \ref{graphs} displays BSEM and BSEMed estimates for the {\em{band}} structure when $n=50$ and $p=100$ using the R package {\em{rags2ridges}} (Peeters and van Wieringen, 2014; van Wieringen and Peeters, 2014). 
Figures \ref{rocs} \& \ref{graphs} show that BSEMed estimates become more accurate when prior knowledge quality increases and are as good as BSEM estimates when using 0\% true edges information.
It is also easy to see (Figure \ref{graphs}) a convergence of the BSEMed estimates to the true graph when the prior knowledge quality increases.							

\section{Illustration}
\label{SectionIllustration}	
We turn to real data in this section. We use gene expression data from the Gene Expression Omnibus (GEO) to illustrate and evaluate methods for reconstructing gene networks. We consider two types of cancer and cancer-related pathways. First,  we focus on the Apoptosis pathway with $p = 84$ genes in a lung data set (Landi et al., 2008), consisting of $n^{\text{lung}}_1=49$ observations from normal tissue and $n^{\text{lung}}_2=58$ observations from tumor tissue, so $n^{\text{lung}}=107$ in total. Secondly, we considered the p53 pathway in a pancreas data set (Badea et al., 2008) with $p = 68$ genes, consisting of
$n^{\text{pancreas}}_1=39$  observations from normal tissue and $n^{\text{pancreas}}_2=39$ observations from tumor tissue, hence $n^{\text{pancreas}}=78$ in total.
Note that the data were scaled per gene prior to the computations.\\
BSEMed and several competing methods (including BSEM) were applied on the tumor data parts of the data sets. For BSEMed, the corresponding data parts from normal tissue were used as prior knowledge by fitting genes networks on the normal data using BSEM. The idea is that, while tumors and normal tissue may differ quite strongly in terms of mean gene expression, the gene-gene interaction network may be relatively more stable. 

\begin{figure}
	\subfigure[BSEM network estimate]{\includegraphics[width=0.45\textwidth]{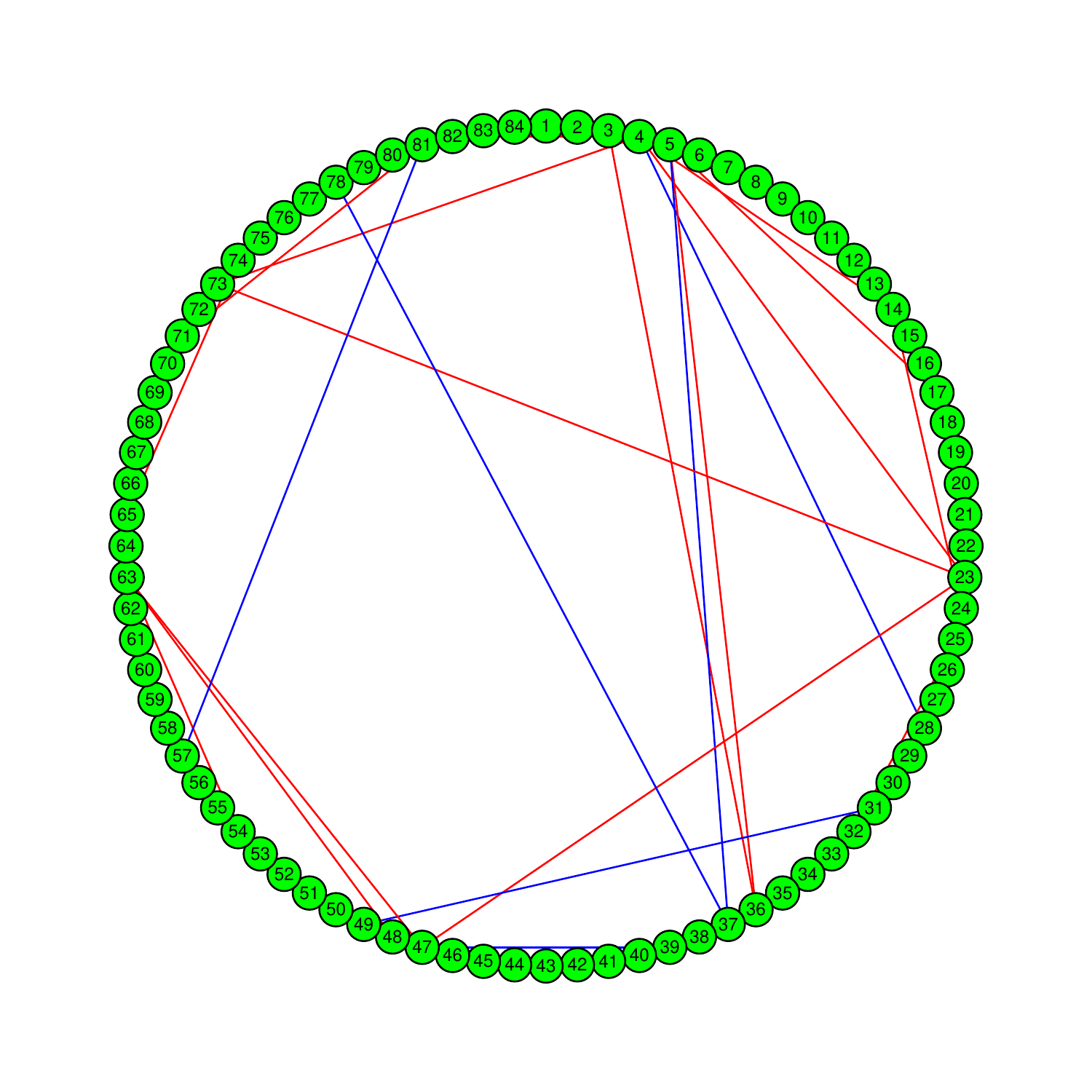}}
	\subfigure[BSEMed network estimate]{\includegraphics[width=0.45\textwidth]{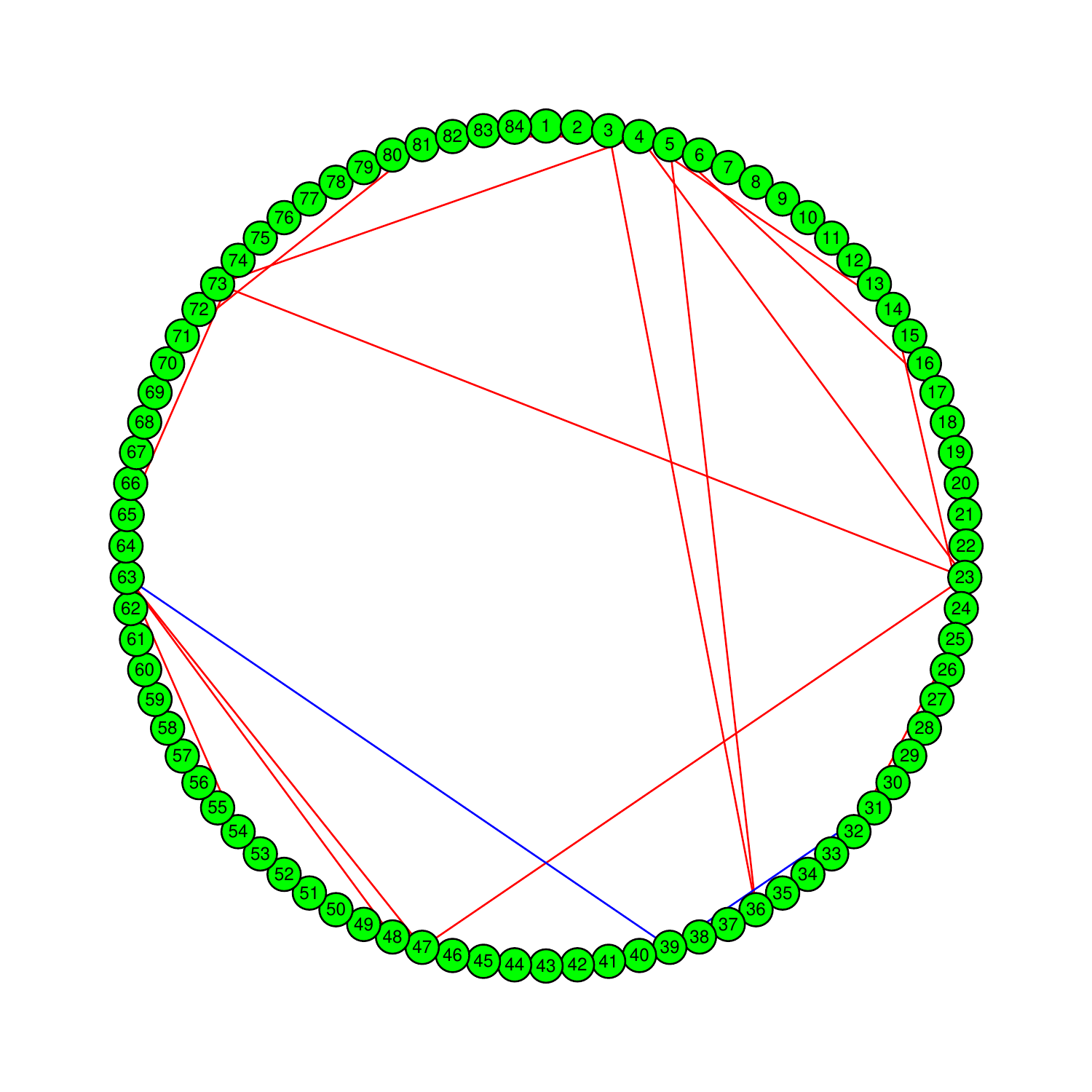}}
	\caption{BSEM vs BSEMed network estimates in lung cancer. Red edges are the overlap edges.}\label{LungNet}
\end{figure}

\begin{figure}[]
	\subfigure[BSEM network estimate]{\includegraphics[width=0.45\textwidth]{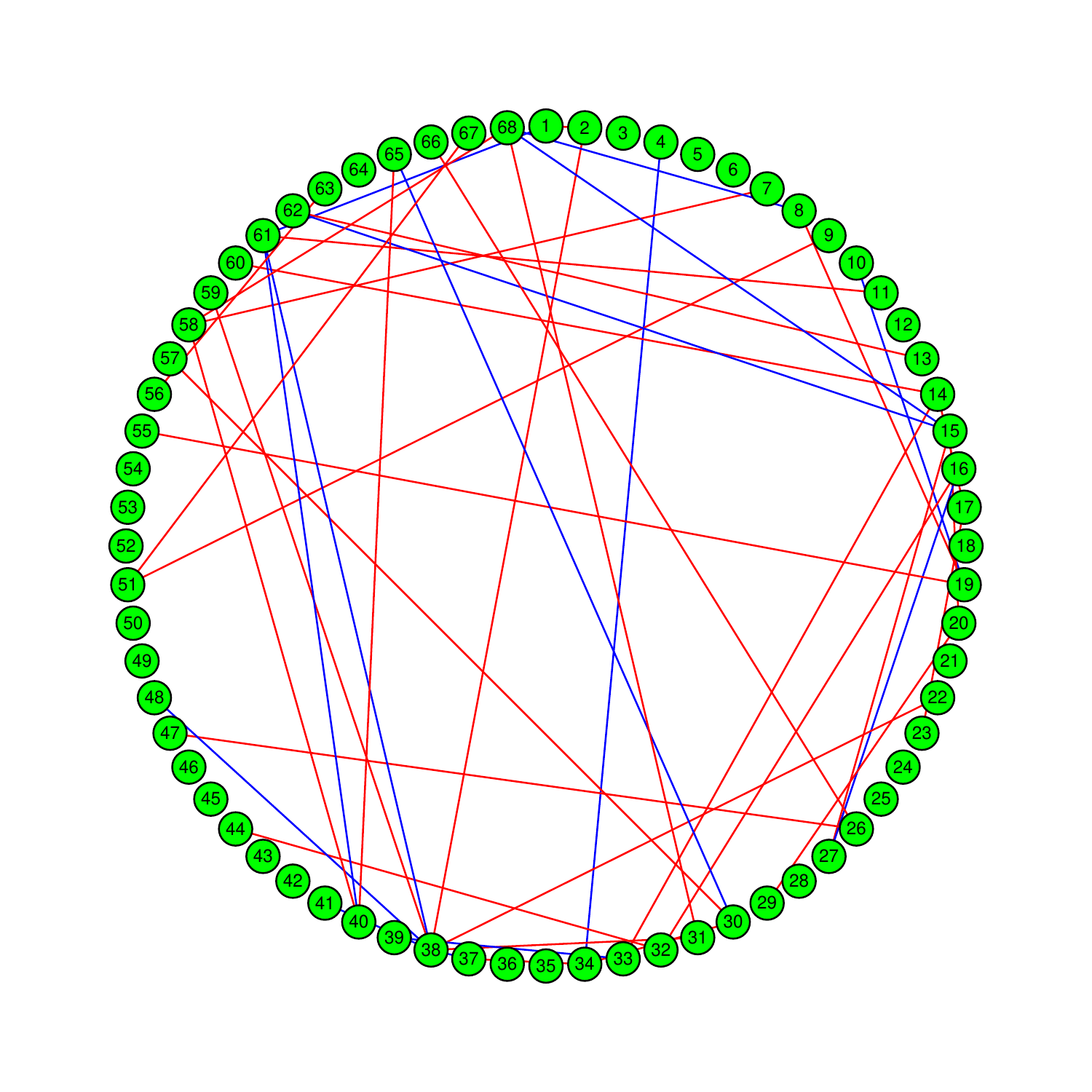}}
	\subfigure[BSEMed network estimate]{\includegraphics[width=0.45\textwidth]{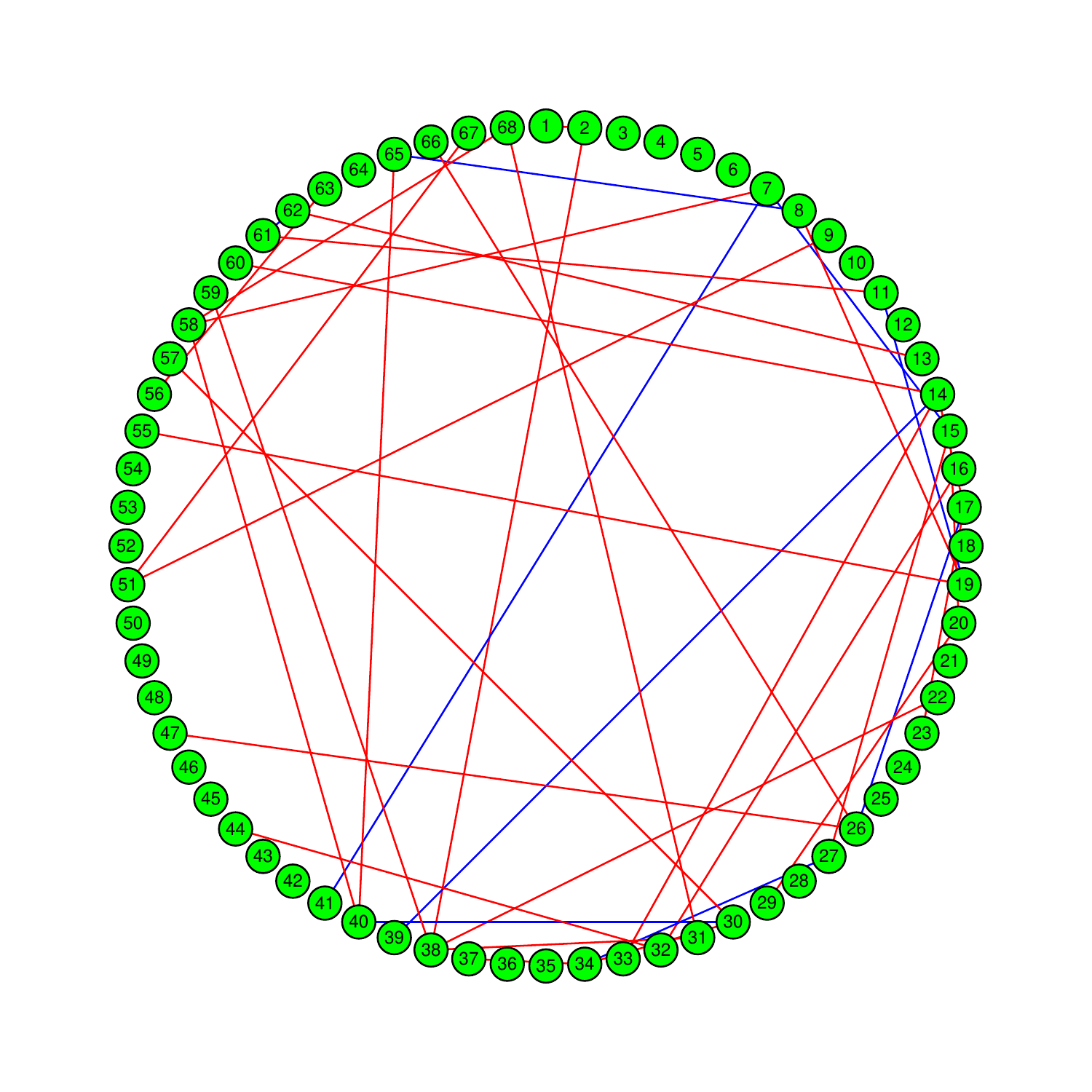}}
	\caption{BSEM vs BSEMed network estimates in pancreas cancer. Red edges are the overlap edges.}\label{PancreasNet}
\end{figure}

We first illustrate the results from BSEM and BSEMed. Figure \ref{LungNet} (Figure \ref{PancreasNet}) displays the estimated gene-gene network interaction in lung cancer (pancreas cancer) and their overlaps. 
Considerable overlap (red edges), but also notable differences can be seen.

Table \ref{EBestimatesData} displays the prior means of precision, as estimated by EB. The prior network is clearly of use: 
the mean prior precision for regression parameters corresponding to the edges absent in the prior network is relatively large, which effectuates stronger shrinkage towards zero than for parameters corresponding to edges present in the prior network.

\begin{center}
	\begin{tabular}{|c | c | c | c|}
		\hline
		& ${\hat{a}_0}/{\hat{b}_0}$ &  ${\hat{a}_1}/{\hat{b}_1}$  &  ratio \\  \hline
		Lung         &   27.32      &    1.71   &   15.97       \\ \hline
		Pancreas   &    20.03     &     1.21  &   12.97          \\
		\hline
	\end{tabular}
	\captionof{table}{EB estimates of precisions $\tau_{i,0}^2$ and $\tau_{i,1}^2$ of prior distributions in lung data (resp. pancreas data) set.}\label{EBestimatesData} 
\end{center}

In the following, we argue that BSEMed network estimates may be more reliable in this setting than those of BSEM, Graphical Lasso (GL$_{\lambda}$) (Friedman et al., 2008), SEM with the Lasso penalty (SEM$_L$) (Meinshausen and B\"uhlmann, 2006) and GeneNet (Sch\"afer et al., 2006) (see the Supplementary Material for methodological details). For that, we assess performance of all methods by studying reproducibility of edges. We randomly split the tumor data part of the lung data set (pancreas data set) into two equal and independent parts: $n^{\text{lung}}_{2,1}$ and $n^{\text{lung}}_{2,2}$ (resp. $n^{\text{pancreas}}_{2,1}$ and $n^{\text{pancreas}}_{2,2}$). BSEM, BSEMed, GL$_{\lambda}$, GeneNet and SEM$_L$ were applied on each subset of the tumor data. We repeated the procedure 50 times. We report in Table \ref{LungRepro} (Table \ref{PancreasRepro}) the average number of overlapping edges between the two subsets for each method when the total number of edges selected by each method on each subset is set to $50$, $100$ and $200$.

\begin{table}[h!]
	\begin{center}
		\begin{tabular}{|c | c | c | c| c|c|c|}
			\hline
			\# edges  &     \vtop{\hbox{\strut BSEM}\hbox{\strut overlap}}  & \vtop{\hbox{\strut GeneNet}\hbox{\strut overlap}} & \vtop{\hbox{\strut SEM$_L$}\hbox{\strut overlap}} &  \vtop{\hbox{\strut GL$_{\lambda}$}\hbox{\strut overlap}}  &  \vtop{\hbox{\strut BSEMed}\hbox{\strut overlap}} &  \vtop{\hbox{\strut \# prior edges}\hbox{\strut in BSEMed}} \\  \hline
			50                          &   4.56      &    1.88      &       1.32   &    3.42      &   29.58  &  13.4  \\ \hline
			100                        &   10.68    &   5.7       &  5.64        &  7.86        &  37.88  &  22.14 \\    \hline
			200                        &   24.16    &   17.2     &  16.46      &  18.14    &  51.54  & 33.7 \\
			\hline
		\end{tabular}
		\captionof{table}{Lung data, reproducibility study: Average number of overlapping edges among the top 50 (100, 200) strongest ones in two equally-sized splits of the tumor data for BSEMed, BSEM, GL$_{\lambda}$, GeneNet and SEM$_L$. }\label{LungRepro}
	\end{center}
\end{table}

\begin{table}[h]
	\begin{center}
		\begin{tabular}{|c | c | c | c| c|c|c|}
			\hline
			\# edges &     \vtop{\hbox{\strut BSEM}\hbox{\strut overlap}}  & \vtop{\hbox{\strut GeneNet}\hbox{\strut overlap}} & \vtop{\hbox{\strut SEM$_L$}\hbox{\strut overlap}} &  \vtop{\hbox{\strut GL$_{\lambda}$}\hbox{\strut overlap}}  &  \vtop{\hbox{\strut BSEMed}\hbox{\strut overlap}} &  \vtop{\hbox{\strut \# prior edges}\hbox{\strut in BSEMed}} \\  \hline
			50                         &   7.42 &  3.32      &  2.8         &      4.52          &   27.82 &  11.92   \\ \hline
			100                       &   17.46  &  10.34         &     9.08     &   11.4  & 57.18 &  29.22   \\    \hline
			200                       & 44.14          &   30.94         &     28.54          &   33.66            &   81.66 & 54.1 \\
			\hline
		\end{tabular}
		\captionof{table}{Pancreas data, reproducibility study: Average number of overlapping edges among the top 50 (100, 200) strongest ones in two equally-sized splits of the tumor data for BSEMed, BSEM, GL$_{\lambda}$, GeneNet and SEM$_L$. }\label{PancreasRepro}
	\end{center}
\end{table}

\begin{figure}[]
	\subfigure[Lung data]{\includegraphics[width=0.45\textwidth]{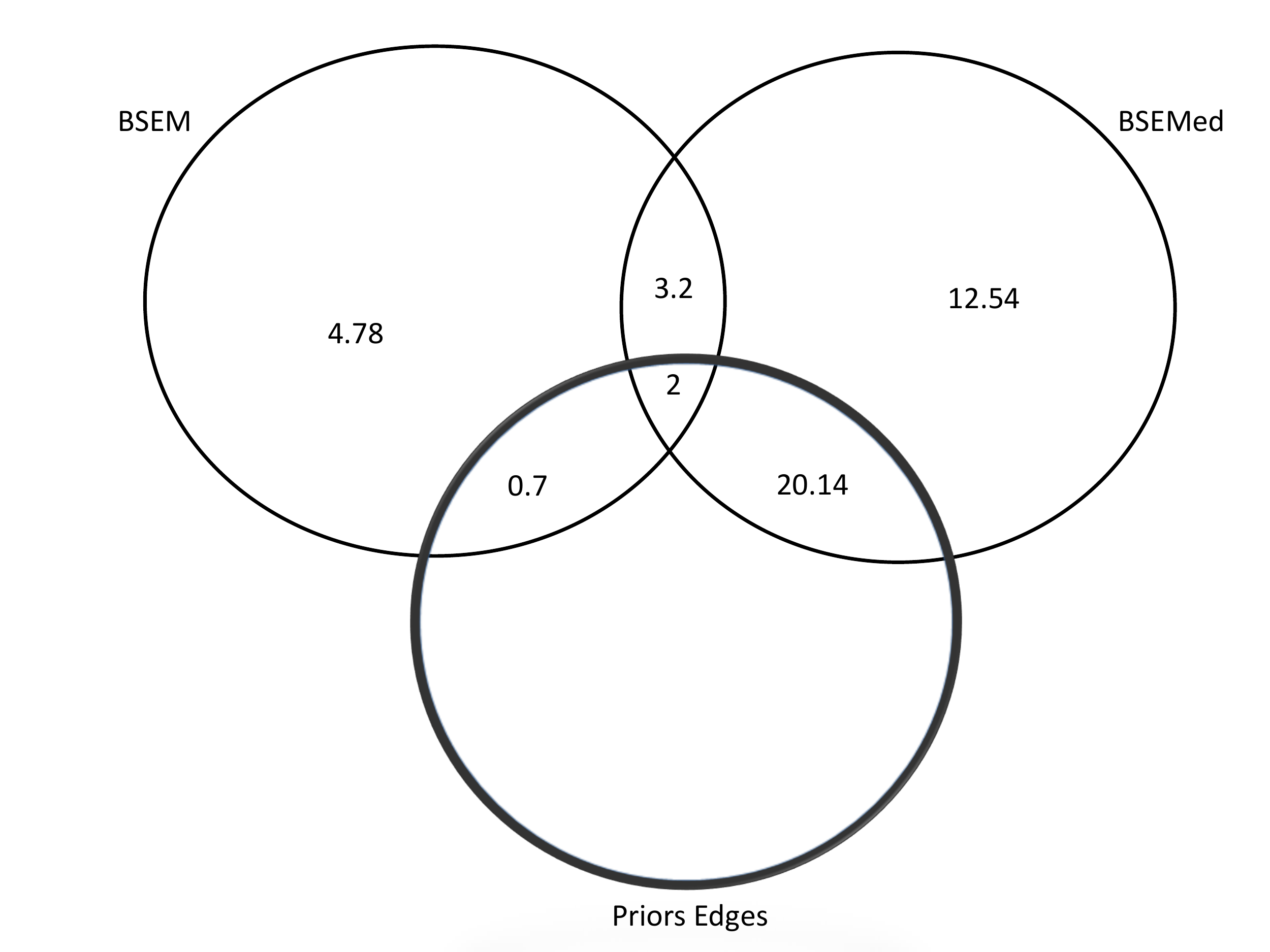}}
	\subfigure[Pancreas data]{\includegraphics[width=0.45\textwidth]{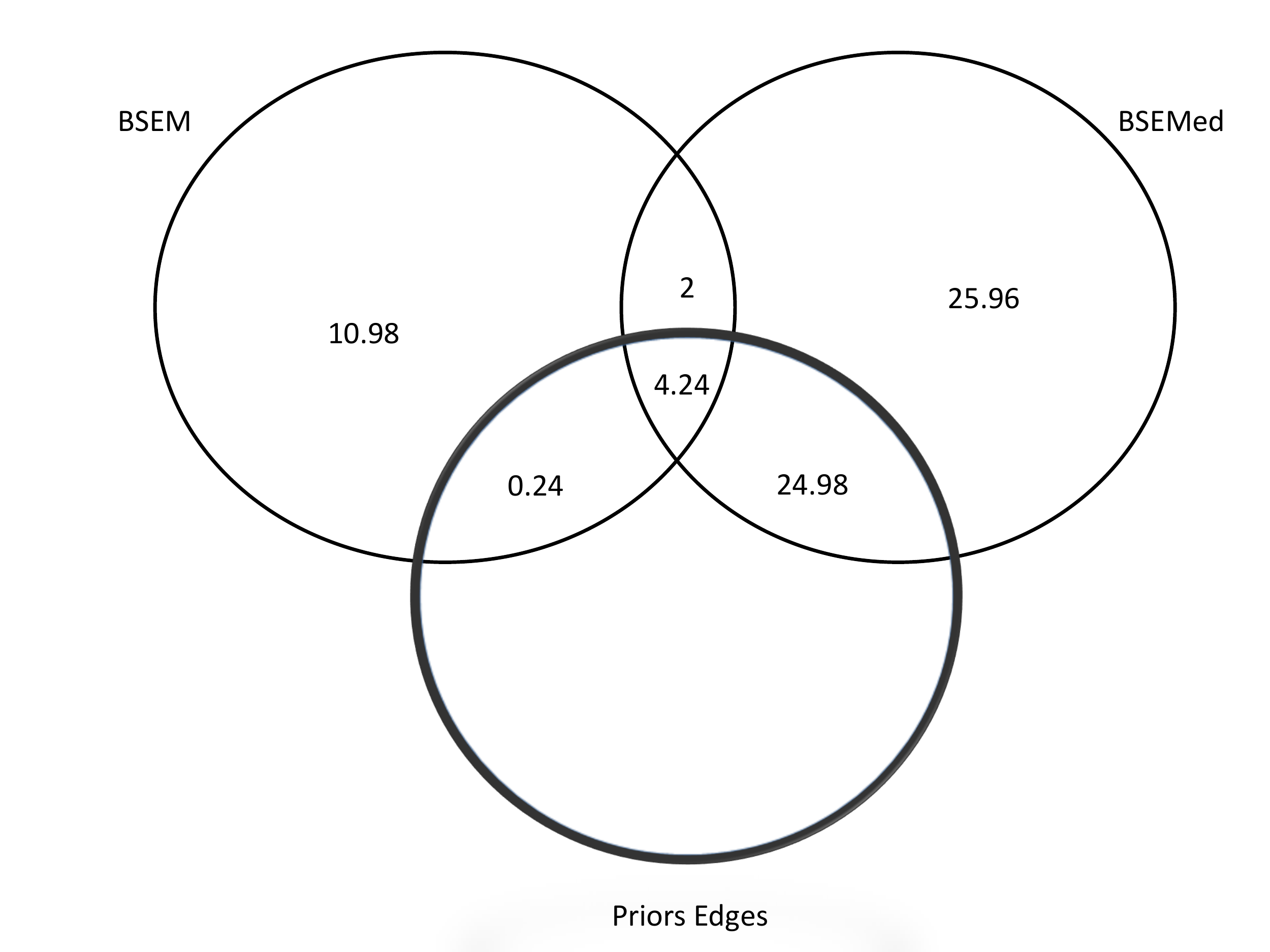}}
	\caption{Venn diagrams displaying  the mean overlap of reproduced top-ranking edges, corresponding to the second row of Table \ref{LungRepro}  (Figure \ref{VennDiagram}.a) and Table \ref{PancreasRepro} (Figure \ref{VennDiagram}.b). }\label{VennDiagram}
\end{figure}

We observe from Tables \ref{LungRepro} \& \ref{PancreasRepro} that the results from the BSEMed networks are much more reproducible than that of BSEM, which is on its turn more reproducible than the other ones. 
Clearly, the improvement can partly be explained by overlapping edges that were also part of the prior network.  However,  it is clear from Figure \ref{VennDiagram} that the BSEMed network estimate in tumor tissue is not just a `finger print' of the prior network (normal tissue network): BSEMed can even reveal edges that are neither in prior network nor in BSEM network estimate.

\begin{figure}[h!]
	\subfigure[Network estimate in Normal tissue]{\includegraphics[width=0.45\textwidth]{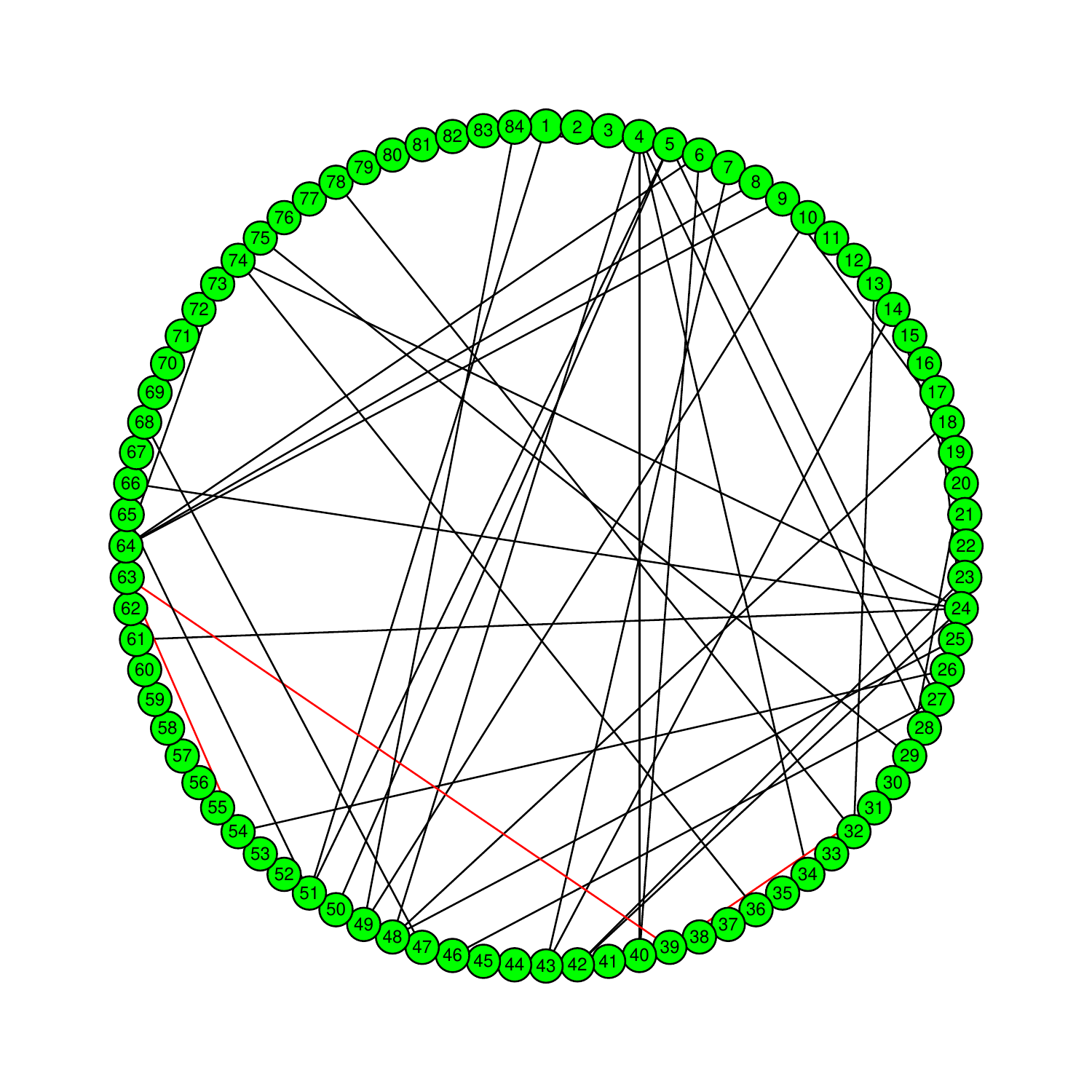}}
	\subfigure[BSEMed network estimate in tumor tissue]{\includegraphics[width=0.45\textwidth]{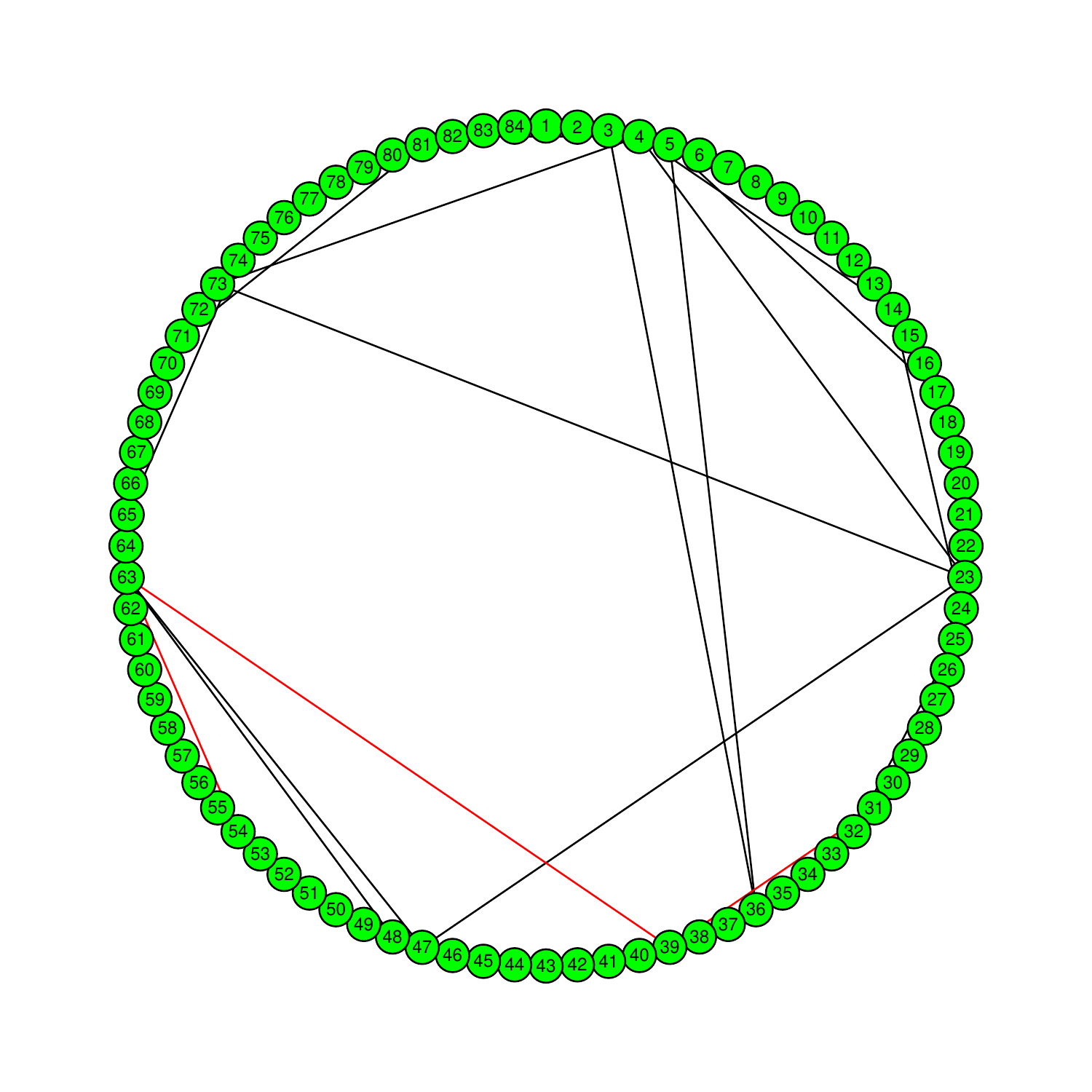}}
	\caption{Network in a normal cell vs BSEMed network in lung cancer. Red edges are the overlap edges between prior and posterior networks.}\label{LungPriorPost}
\end{figure}

\begin{figure}[h]
	\subfigure[Network estimate in Normal tissue]{\includegraphics[width=0.45\textwidth]{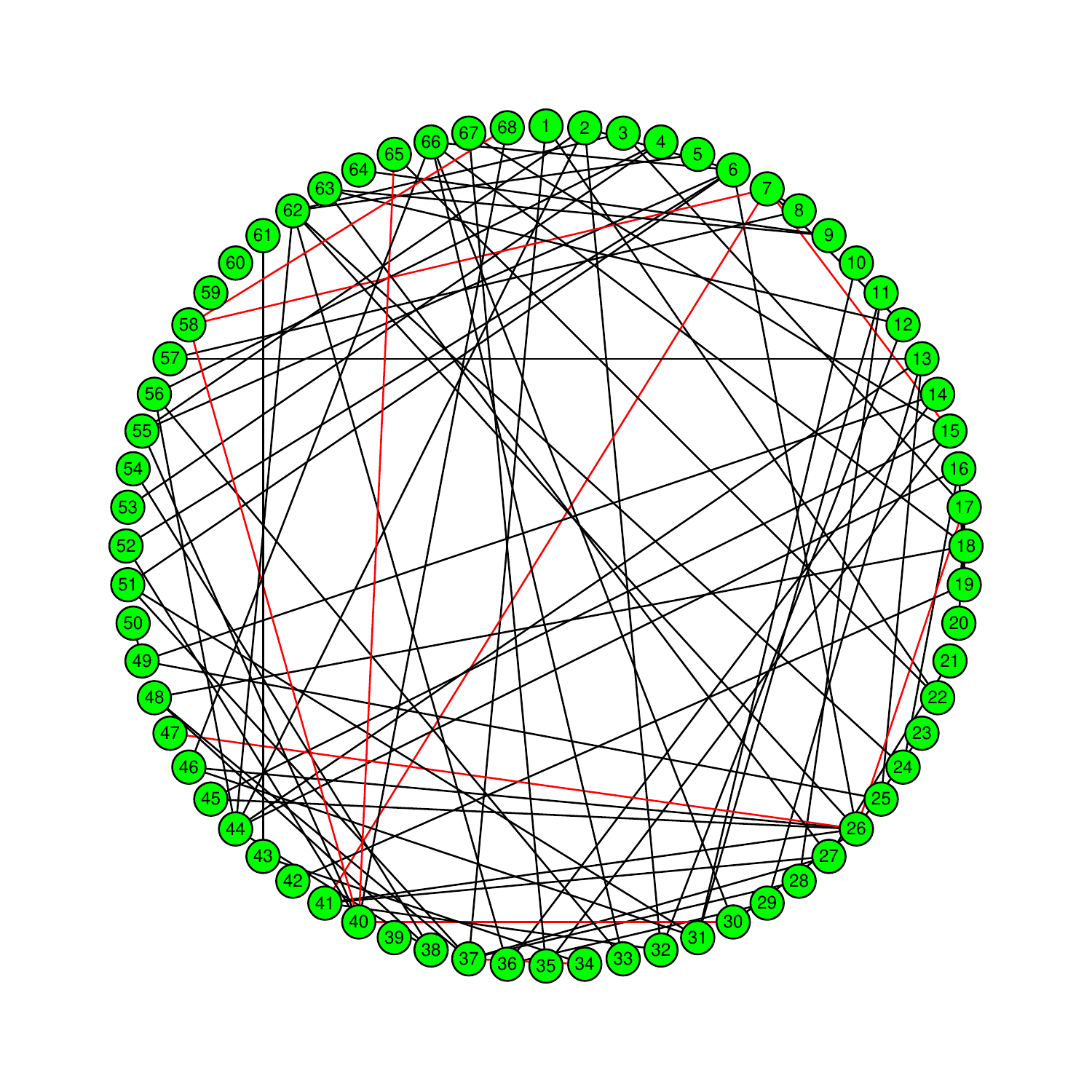}}		\subfigure[BSEMed network estimate in tumor tissue]{\includegraphics[width=0.45\textwidth]{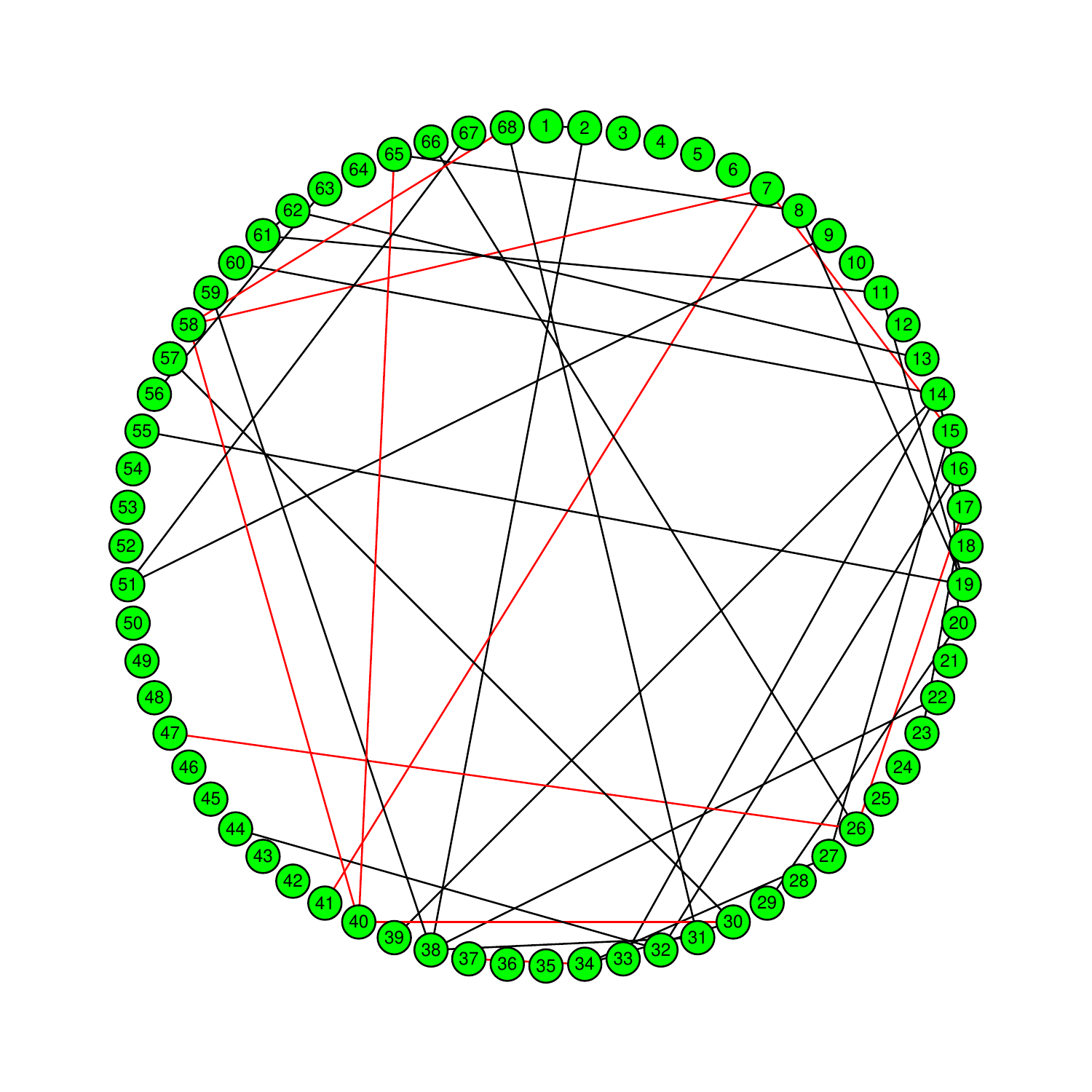}}
	\caption{Network in a normal cell vs BSEMed network in pancreas cancer. Red edges are the overlap edges between prior and posterior networks.}\label{PancreasPriorPost}
\end{figure}

Figure \ref{LungPriorPost} (resp. Figure \ref{PancreasPriorPost}) displays the network in normal tissue against the network in tumor tissue in the lung data (resp. in the pancreas data).


\clearpage

\section{Discussion}

We have presented a new method for incorporating prior information in undirected network reconstrustion based on Bayesian SEM. Our approach allows the use of two  central Gaussian distributions per regression equation for coefficients $\beta_{i,r}$'s of our SEMs, where the prior information determines which of the two applies to a specific $\beta_{i,r}$. Empirical bayes estimation of the parameters of the two hyper priors of the precisions effectuates shrinkage and accommodates the situation where the prior information would not be relevant. We showed in simulation with different graph structures that BSEMed performs clearly better than BSEM when the used prior knowledge is relevant and as good as when not. In addition, for two real data sets we showed better reproducibility of top ranking edges with respect to other methods .

Instead of assigning Gaussian distributions to the coefficients, other (e.g. sparse) priors can be used. However, the fast variational Bayes method for posterior density approximation may not be of use anymore. For instance, would one use  Horseshoe priors (Carvalho et al., 2010), the variational marginals are non-existent. 
The complement property (Section \ref{SectionModel} )
is preserved whenever the same functional forms of the priors are used for both classes. However, a combination of e.g. a Gaussian and a
sparse prior ruins this property, which renders such a combination less attractive.

Future research also focuses on extending our method to situations with more than
two classes. For example, when considering integrative networks for two sets of molecular markers or two (related) pathways,
the three class setting is relevant: two classes represent the connections within the two sets and a third one between the two sets.
Finally, multiple sources of external data may be at one's disposal and need incorporation in BSEMed. This requires to model the parameter(s) of the priors in terms of contibutions of those external sources, and weigh those sources in a data-driven manner, as it is unlikely that the sources are equally informative.

\end{document}